\documentclass[amsmath,amssymb,prb,superscriptaddress,showpacs,longbibliography,onecolumn]{revtex4-1}
\usepackage{graphicx}
\usepackage{multirow}
\usepackage{upgreek}
\usepackage{color}

\usepackage{eepic,epic}
\usepackage[colorlinks=true,linkcolor=black,citecolor=black,urlcolor=blue]{hyperref}
\usepackage{stmaryrd}
\usepackage{dsfont}
\usepackage{esvect}
\usepackage{lipsum}
\usepackage{xr}
\usepackage{upgreek}
\usepackage{enumitem}

\usepackage{tikz}
\usetikzlibrary{patterns}
\usetikzlibrary{decorations.markings}
\usetikzlibrary{positioning}

\newcommand\dd{\mathrm{d}}
\newcommand\ud{\mathrm{d}}
\newcommand\Tr{\mathrm{Tr}\,}
\newcommand\im{\mathrm{Im} \ }

\newcommand\iu{\mathrm{i}\,}

\usepackage{siunitx}
\DeclareSIUnit\mub{\mu_\mathrm{B}}
\DeclareSIUnit\rydberg{\mathrm{Ry}}
\usepackage{mathtools}
\DeclarePairedDelimiter\bra{\langle}{\rvert}
\DeclarePairedDelimiter\ket{\lvert}{\rangle}
\DeclarePairedDelimiterX\braket[1]{\langle}{\rangle}{#1}
\renewcommand{\vec}[1]{\mathbf{#1}}

\begin{document}
	
	\tikzset{
		every node/.append style={font=\small},
		every edge/.append style={thick},
		arrow/.style={thick, shorten >=5pt,shorten <=5pt,->},
		Green_soc/.style={dashed, draw, postaction={decorate},
			decoration={markings,mark=at position .55 with {\arrow[draw]{>}}}},
		Green_full/.style={solid, double, draw, postaction={decorate},
			decoration={markings,mark=at position .55 with {\arrow[draw]{>}}}},
		Green/.style={solid, draw, postaction={decorate},
			decoration={markings,mark=at position .55 with {\arrow[draw]{>}}}},
		susc/.style={thick,fill=white,draw},
		dt/.style={thick,fill=lightgray,draw},
		soc/.style={thick,fill=black,draw},
		vertex/.style={thick,fill=black,draw}
	}

	\title{Generalization of the Landau-Lifshitz-Gilbert equation by multi-body contributions to Gilbert damping for non-collinear magnets}
	
	\author{Sascha Brinker}
	\affiliation{Peter Gr\"{u}nberg Institut and Institute for Advanced Simulation, Forschungszentrum J\"{u}lich \& JARA, 52425 J\"{u}lich, Germany}
	\author{Manuel dos Santos Dias}\email{m.dos.santos.dias@fz-juelich.de}
	\affiliation{Faculty of Physics, University of Duisburg-Essen and CENIDE, 47053 Duisburg, Germany}
	\affiliation{Peter Gr\"{u}nberg Institut and Institute for Advanced Simulation, Forschungszentrum J\"{u}lich \& JARA, 52425 J\"{u}lich, Germany}
	\author{Samir Lounis}\email{s.lounis@fz-juelich.de}
	\affiliation{Peter Gr\"{u}nberg Institut and Institute for Advanced Simulation, Forschungszentrum J\"{u}lich \& JARA, 52425 J\"{u}lich, Germany}
	\affiliation{Faculty of Physics, University of Duisburg-Essen and CENIDE, 47053 Duisburg, Germany}

	\date{\today}

	
	\begin{abstract}
		\noindent
		We propose a systematic and sequential expansion of the Landau-Lifshitz-Gilbert equation utilizing the dependence of the Gilbert damping tensor on the angle between magnetic moments, which arises from multi-body scattering processes. The tensor consists of a damping-like term and a correction to the gyromagnetic ratio. Based on electronic structure theory, both terms are shown to depend on e.g. the scalar, anisotropic, vector-chiral and scalar-chiral products of magnetic moments: $\vec{e}_i\cdot\vec{e}_j$, $(\vec{n}_{ij}\cdot\vec{e}_i)(\vec{n}_{ij}\cdot\vec{e}_j)$,  $\vec{n}_{ij}\cdot(\vec{e}_i\times\vec{e}_j)$, 
		$(\vec{e}_i\cdot\vec{e}_j)^2$,  
		$\vec{e}_i\cdot(\vec{e}_j\times\vec{e}_k)$..., where some terms are subjected to the spin-orbit field $\vec{n}_{ij}$ in  first and second order. We explore the magnitude of the different contributions using both the Alexander-Anderson model and time-dependent density functional theory in magnetic adatoms and dimers deposited on Au(111) surface.
	\end{abstract}
	
	
	\maketitle

	
	\section{Introduction}
	In the last decades non-collinear magnetic textures have been at the forefront in the field of spintronics due to the promising applications and perspectives tied to them~\cite{Fert2013,Fert2017}. 
	Highly non-collinear particle-like topological swirls, like skyrmions~\cite{Bogdanov1994,Roessler2006} and hopfions~\cite{Tai2018}, but also domain walls~\cite{Parkin2008} can potentially be utilized in data storage and processing devices with superior properties compared to conventional devices. Any manipulation, writing and nucleation of these various magnetic states involve magnetization dynamical processes, which are crucial to understand for the design of future spintronic devices. 
	
	In this context, the Landau-Lifshitz-Gilbert (LLG) model~\cite{Landau1935,Gilbert2004} is widely used to describe spin dynamics of materials ranging from 3-dimensional bulk magnets down to the 0-dimensional case of single atoms, see e.g. Refs.~\cite{Eriksson2017,dos_santos_dias_relativistic_2015,Lounis2015,Guimaraes2017}.
	The LLG model has two important ingredients: (i) the Gilbert damping being in general a tensorial quantity~\cite{Bhattacharjee2012}, which can originate from the presence of spin-orbit coupling (SOC)~\cite{Kambersky1970} and/or from spin currents pumped into a reservoir~\cite{Mizukami2002,Tserkovnyak2002}; (ii) the effective magnetic field acting on a given magnetic moment and rising from internal and external interactions. 
	
	Often a generalized Heisenberg model, including magnetic anisotropies and magnetic exchange interactions, is utilized to explore the ground state and magnetization dynamics characterizing a material of interest. Instead of the conventional bilinear form, the magnetic interactions can eventually be of higher-order type,   see e.g.~\cite{Hayami2017,Brinker2019,Laszloffy2019,Grytsiuk2020,Brinker2020,Lounis2020,Dias2022}. Similarly to magnetic interactions, the Gilbert damping, as we demonstrate in this paper, can host higher-order non-local contributions. Previously, signatures of giant anisotropic damping were found~\cite{Jue2015}, while chiral damping and renormalization of the  gyromagnetic ratio were revealed through measurements executed on chiral domain wall creep motion~\cite{Jue2015,Akosa2016,Freimuth2017,Akosa2018,Kim2018}.
	
	Most first-principles studies of the Gilbert damping were either focusing on collinear systems or were case-by-case studies on specific non-collinear structures lacking a general understanding of the fundamental behaviour of the Gilbert damping as function of  the non-collinear state of the system. 
	In this paper, we discuss the Gilbert damping tensor and its dependencies on the alignment of spin moments as they occur in  arbitrary non-collinear state. Utilizing linear response theory, we extract the dynamical magnetic susceptibility and identify the Gilbert damping tensor pertaining to the generalized LLG equation that we map to that obtained from  electronic structure models such as the single orbital Alexander-Anderson model~\cite{Alexander1964} or time-dependent density functional theory applied to realistic systems~\cite{Gross1985,Lounis2010,dos_santos_dias_relativistic_2015}.  
	Applying systematic perturbative expansions, we find the allowed dependencies of the Gilbert damping tensor on the direction of the magnetic moments. We identify terms that are affected by SOC in first and second order.
	We generalize the LLG equation by a simple form where the Gilbert damping tensor is amended with terms proportional to scalar, anisotropic, vector-chiral and scalar-chiral products of magnetic moments, i.e. terms like $\vec{e}_i\cdot\vec{e_j}$, $(\vec{n}_{ij}\cdot\vec{e}_i)(\vec{n}_{ij}\cdot\vec{e}_j)$, $\vec{n}_{ij}\cdot(\vec{e}_i\times\vec{e_j})$, 
	$(\vec{e}_i\cdot\vec{e_j})^2$, $\vec{e}_i\cdot(\vec{e_j}\times\vec{e_k})$..., where we use unit vectors, $\vec{e}_i = \vec{m}_i / |\vec{m}_i|$, to describe the directional dependence of the damping parameters and $\vec{n}_{ij}$ represents the spin-orbit field.
	
	The knowledge gained from the Alexander-Anderson model is applied to realistic systems obtained from first-principles calculations.
	As prototypical test system we use $3d$ transition metal adatoms and dimers deposited on the Au(111) surface.
	Besides the intra-site contribution to the Gilbert damping, we also shed light on the inter-site contribution, usually referred to as the non-local contribution.
	
	\begin{figure*}[!tb]
		\begin{center}
			\includegraphics[width=\textwidth
			]{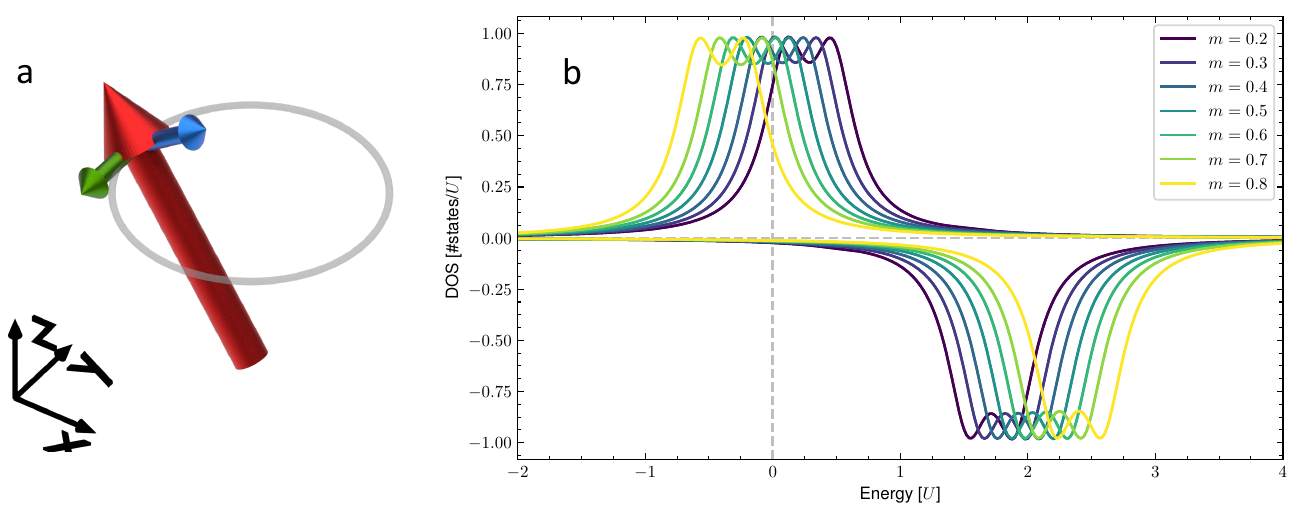}
			\caption{\label{fig:figure1}
				Illustration of the Landau-Lifshitz-Gilbert model and local density of states within the Alexander-Anderson model. (a)
				A magnetic moment (red arrow) precesses in the the presence of an external field.
				The blue arrow indicates the direction of a damping term, while the green arrow shows the direction of the precession term. (b) Density of states for different magnetizations in the range from $\SI{0.2}{}$ to $\SI{0.8}{}$.
				Density of states of dimers described within the Alexander-Anderson model for different magnetizations in the range from $\SI{0.2}{}$ to $\SI{0.8}{}$.
				Shown is the ferromagnetic reference state.
				The magnetizations are self-consistently constrained using a longitudinal magnetic field, which is shown in the inset.
				Model parameters: $U=\SI{1.0}{\electronvolt}$, $E_d = \SI{1.0}{\electronvolt}, \ t = \SI{0.2}{\electronvolt}, \ \Gamma = \SI{0.2}{\electronvolt} \ , \ \varphi_\text{R} = \SI{0}{\degree}$. 
			}
		\end{center}
	\end{figure*}

	\section{Mapping the Gilbert damping from the dynamical magnetic susceptibility}
	Here we extract the dynamical transverse magnetic response of a magnetic moment from both
	the Landau-Lifshitz-Gilbert model and electronic structure theory in order to identify the Gilbert damping tensor $\mathcal{G}_{ij}$~\cite{Lounis2011,dos_santos_dias_relativistic_2015,Lounis2015,Guimaraes2019}. 
	In linear response theory, the response of the   magnetization $\vec{m}$ at site $i$ to a transverse magnetic field $\vec{b}$ applied at sites $j$ and oscillating at frequency $\omega$ reads
	\begin{align}
		m^\alpha_i(\omega) = \sum_{j\beta}  \chi_{ij}^{\alpha \beta}(\omega) b_j^\beta(\omega) \quad ,
	\end{align}
	with the magnetic susceptibility $\chi_{ij}^{\alpha \beta}(\omega)$ and $\alpha,\beta$ are the $x,y$ coordinates defined in the local spin frame of reference pertaining to sites $i$ and $j$.
	
	In a general form \cite{Bhattacharjee2012} the LLG equation is given by
	\begin{align}
		\frac{\dd \vec{m}_i}{\dd t} = - \gamma \vec{m}_i \times \left( \vec{B}_i^\text{eff} + \sum_j \mathcal{G}_{ij} \cdot \frac{\dd \vec{m}_j}{\dd t} \right) \quad , \label{eq:general_LLG}
	\end{align}
	where $\gamma = 2$ is the gyromagnetic ratio, $\vec{B}_i^\text{eff}= - \dd \mathcal{H}^\text{spin} / \dd \vec{m}_i$ is the effective magnetic field
	containing the contributions from an external magnetic field $\vec{B}_i^\text{ext}$, as well as internal magnetic fields originating from the interaction of the moment with its surrounding.
	In an atomistic spin model described by e.g.~the generalized Heisenberg hamiltonian, $
	\mathcal{H}^\text{spin} = \sum_{i} \vec{m}_i \, K_i \, \vec{m}_i + \frac{1}{2} \sum_{ij} \vec{m}_i \, J_{ij} \, \vec{m}_j$,
	containing the on-site magnetic anisotropy $K_i$ and the exchange tensor $J_{ij}$, the effective field is given by $\vec{B}_i^\text{eff} = \vec{B}_i^\text{ext} - K_i \, \vec{m}_i - \sum_j J_{ij} \, \vec{m}_j$ (green arrow in Fig.~1a). 
	The Gilbert damping tensor can be separated into two contributions -- a damping-like term, which is the symmetric part of the tensor, $\mathcal{S}$, (blue arrow in Fig.~1a), and a precession-like term $\mathcal{A}$, which is the anti-symmetric part of the tensor.
	In Appendix~\ref{app:Gilbert_damping_tensor} we show how the antisymmetric intra-site part of the tensor contributes to a renormalization of the gyromagnetic ratio.

	To extract the magnetic susceptibility, we express the magnetic moments in their respective local spin frame of references and use Rotation matrices that ensure rotation from local to globa spin frame of reference (see Appendix~\ref{app:relation_LLG_susc}). 
	The magnetic moment is assumed to be perturbed around its equilibrium value $M_i$, $ \vec{m}_i^\text{loc} = M_i \vec{e}_i^z + m_i^x \vec{e}_i^x  + m_i^y \vec{e}_i^y $, where $\vec{e}_i^\alpha$ is the unit vector in direction $\alpha$ in the local frame of site $i$. Using the ground-state condition of vanishing magnetic torques, $M_i \vec{e}_{i}^z \times \left( \vec{B}_i^\text{ext} + \vec{B}_i^\text{int} \right) = 0$ and the inverse of the transverse magnetic susceptibility can be identified as 
		\begin{align}
			\chi^{-1}_{i\alpha j\beta}(\omega) = \delta_{ij} \left( \delta_{\alpha \beta} \frac{B_{iz}^\text{eff}}{M_i} + \frac{\iu \, \omega}{\gamma M_i} \epsilon_{\alpha \beta \mu} \right)
			+ \frac{1}{M_i M_j}( \mathcal{R}_i J_{ij} \mathcal{R}_j^\mathsf{T} )_{\alpha \beta} + \iu \, \omega ( \mathcal{R}_i \mathcal{G}_{ij} \mathcal{R}_j^\mathsf{T} )_{\alpha \beta} \quad , \label{eq:susc_LLG_relation}
	\end{align}
	from which it follows that the Gilbert damping is directly related to the linear in frequency imaginary part of the inverse susceptibility
	\begin{align}
		\frac{\dd}{\dd \omega} \Im [\chi^{-1}]_{ij}^{\alpha \beta}= \delta_{ij} \left( \frac{1}{\gamma M_i} \epsilon_{\alpha \beta \mu} \right)
		+  ( \mathcal{R}_i \mathcal{G}_{ij} \mathcal{R}_j^\mathsf{T} )_{\alpha \beta} \quad . \label{eq:susc_relation_gilbert_damping}
	\end{align}
	Note that $\mathcal{R}_i$ and $\mathcal{R}_j$ are rotation matrices rotating to the local frames of site $i$ and $j$, respectively, which define the coordinates $\alpha, \beta = \{x,y\}$ (see Appendix~\ref{app:relation_LLG_susc}).
	
	Based on electronic structure theory, the transverse dynamical susceptibility can be extracted from a Dyson-like equation: $\chi^{-1}(\omega) = \chi_0^{-1}(\omega) -U $, where $\chi_0$ is the susceptibility of non-interacting electron while $U$ is a many-body interaction Kernel, called exchange-correlation Kernel in the context of time-dependent density functional theory~\cite{Gross1985}. The Kernel is generally assumed to be adiabatic, which enables the evaluation of the Gilbert damping directly from the non-interacting susceptibility. Obviously: $\frac{\dd}{\dd \omega}\chi^{-1}(\omega) = \frac{\dd}{\dd \omega}\chi^{-1}_0(\omega)$. For small frequencies $\omega$, $\chi_0$ has a simple $\omega$-dependence~\cite{Lounis2015}:
	\begin{align}
		\chi_0 (\omega) \approx \Re \chi_0 (0) + i \omega \Im \frac{\dd}{\dd \omega}\chi_0|_{\omega=0}
	\end{align}
	and as shown in Ref.~\cite{Guimaraes2019}
	\begin{align}
		\frac{\dd}{\dd \omega}\chi_0^{-1} (\omega) \approx [\Re \chi_0 (0)]^{-2}   \Im \frac{\dd}{\dd \omega}\chi_0|_{\omega=0}\;.
	\end{align}
	
	Starting from the electronic Hamiltonian $\mathcal{H}$ and the corresponding Green functions $G(E\pm i\eta) = (E-\mathcal{H}\pm i\eta)^{-1}$, one can show that the non-interacting magnetic susceptibility can be defined via
		\begin{align}
			\chi_{0,ij}^{\alpha\beta}(\omega+\iu\eta) = -\frac{1}{\pi}\,\text{Tr}\!\!\int^{E_F}\hspace{-0.5em}\ud E\;\big[& \sigma^\alpha \,G_{ij}(E+\omega+\iu\eta)\,\sigma^\beta\,\im G_{ji}(E) + \sigma^\alpha\,\im G_{ij}(E)\,\sigma^\beta\,G_{ji}(E-\omega-\iu\eta)\big] \quad , \label{eq:susc_definition_green}
		\end{align}
	with $\boldsymbol{\sigma}$ being the vector of Pauli matrices. Obviously to identify the Gilbert damping and how it reacts to magnetic non-collinearity, we have to inspect the dependence of the susceptibility, and therefore the Green function, on the misalignment of the magnetic moments.

	\section{Multi-site expansion of the Gilbert damping}\label{sec:sys_expansion_Gilbert_damping}

	Assuming the hamiltonian $\mathcal{H}$ consisting of an on-site contribution $\mathcal{H}^0$ and an inter-site term encoded in a hopping term $t$, which can be spin-dependent, one can proceed with a perturbative expansion of the corresponding Green function utilizing the Dyson equation
	\begin{align}
		G_{ij} = G_i^0 \delta_{ij} + G_i^0 t_{ij} G_j^0 + G_i^0 t_{ik} G_k^0 t_{kj} G_j^0 + ... \quad . \label{eq:perturbative_expansion}
	\end{align}
	Within the Alexander-Anderson  single-orbital  impurity model~\cite{Alexander1964}, $\mathcal{H}^0_i=E_d - \iu \Gamma - U_i \vec{m}_i \cdot \vec{\sigma} - \vec{B}_i \cdot \vec{\sigma}$, where $E_d$ is the energy of the localized orbitals, $\Gamma$ is the hybridization in the wide band limit, $U_i$ is the local interaction responsible for the formation of a magnetic moment and  $\vec{B}_i$ is an constraining or external magnetic field. SOC can be incorportated as ${t}^{\text{soc}}_{ij} = i \lambda_{ij} \vec{n}_{ij} \cdot \vec\sigma$, where $\lambda_{ij}$ and $\vec{n}_{ij}=-\vec{n}_{ji}$ represent respectively the strength and direction of the anisotropy field. It can be parameterized as a spin-dependent hopping using the  Rashba-like spin-momentum locking $t_{ij} = t \left(\cos \varphi_\text{R} \, \sigma_0  - \iu  \sin \varphi_\text{R} \, \vec{n}_{ij} \cdot \vec{\sigma}\right)$~\cite{Chaudhary2018}.

	Depending on whether the considered Green function is an on-site Green function $G_{ii}$ or an inter-site Green function $G_{ij}$ different orders in the hopping are relevant.
	On-site Green functions require an even number of hopping processes, while inter-site Green functions require at least one hopping process.
	
	The on-site Green function $G_i^0$ can be separated into a spin-less part $\mathcal{N}_{i}$ and a spin dependent part $\boldsymbol{\mathcal{M}}_{i}$,
	\begin{align}
		G^0_{i} = \mathcal{N}_{i} \, \sigma_0 + \boldsymbol{\mathcal{M}}_{i} \cdot \vec{\sigma} \quad , \label{eq:Green_separation_A_B}
	\end{align}
	where the spin dependent part is parallel to the magnetic moment of site $i$, $\boldsymbol{\mathcal{M}}_{i} \parallel \vec{m}_i$
	 (note that SOC is added later on to the hoppings).
	Using the perturbative expansion, eq.~\eqref{eq:perturbative_expansion}, and the separated Green function, eq.~\eqref{eq:Green_separation_A_B}, to calculate the magnetic susceptibility, eq.~\eqref{eq:susc_definition_green}, one can systematically classify the allowed dependencies of the susceptibility with respect to the directions of the magnetic moments, e.g. by using diagrammatic techniques as shown in Ref.~\cite{Brinker2019} for a related model in the context of higher-order magnetic exchange interactions.
	
	Since our interest is in the form of the Gilbert damping, and therefore also in the form of the magnetic susceptibility, the perturbative expansion can be applied to the magnetic susceptibility.
	The general form of the magnetic susceptibility in terms of the Green function, eq.~\eqref{eq:susc_definition_green}, depends on a combination of two Green functions with different energy arguments, which are labeled as $\omega$ and $0$ in the following.
	The relevant structure is then identified as~\cite{Guimaraes2019},
		\begin{align}
		\chi_{ij}^{\alpha  \beta}(\omega) \sim \Tr \sigma^\alpha_i G_{ij}(\omega) \sigma^\beta_j G_{ji}(0) \quad . \label{eq:susc_starting_point_expansion}
\end{align}
	
	The sake of the perturbative expansion is to gather insights in the possible forms and dependencies on the magnetic moments of the Gilbert damping, and not to calculate explicitly the strength of the Gilbert damping from this expansion.
	Therefore, we focus on the structure of eq.~\eqref{eq:susc_starting_point_expansion}, even though the susceptibility has more ingredients, which are of a similar form.
	
	Instead of writing all the perturbations explicitly, we set up a diagrammatic approach, which has the following ingredients and rules:
	\begin{enumerate}
		\item   Each diagram contains the operators $\mathcal{N}$ and $\mathcal{M}$, which are $\sigma^\alpha$ and $\sigma^\beta$ for the magnetic susceptibility.
		The operators are represented by a white circle with the site and spin index: \begin{tikzpicture}[baseline={([yshift=-.5ex]current bounding box.center)}]
			\begin{scope}[scale=2]
				\draw[susc] (0,0) circle (2pt) node[right,xshift=.2cm] {$i \alpha$} ;
			\end{scope}
		\end{tikzpicture}
		
		\item Hoppings are represented by grey circles indicating the hopping from site $i$ to $j$: \begin{tikzpicture}[baseline={([yshift=-.5ex]current bounding box.center)}]
			\begin{scope}[scale=2]
				\draw[dt] (0,0) circle (2pt) node[right,xshift=.2cm] {$ ij $} ;
			\end{scope}
		\end{tikzpicture}.
		The vertex corresponds to $t_{ij}$.

		\item SOC is described as a spin-dependent hopping from site $i$ to $j$ and represented by: \begin{tikzpicture}[baseline={([yshift=-.5ex]current bounding box.center)}]
			\begin{scope}[scale=2]
				\draw[soc] (0,0) circle (2pt) node[right,xshift=.2cm] {$ ij,\alpha $} ;
			\end{scope}
		\end{tikzpicture}.
		The vertex corresponds to ${t}^{\text{soc}}_{ij} = i \lambda_{ij} \hat{n}_{ij}^\alpha \sigma^\alpha$. 
		
		\item The bare spin-independent (on-site) Green functions are represented by directional lines with an energy attributed to it:
		\begin{tikzpicture}[baseline={([yshift=-.5ex]current bounding box.center)}]
			\begin{scope}[scale=2]
				\draw (0,0) edge[Green] node[above] {$\omega$} (.5,0);
			\end{scope}
		\end{tikzpicture}.
		The Green function connects operators and hoppings.
		The line corresponds to $\mathcal{N}_{i}(\omega)$.

		\item The spin-dependent part of the bare Green function is represented by: \begin{tikzpicture}[baseline={([yshift=-.5ex]current bounding box.center)}]
			\begin{scope}[scale=2]
				\draw (0,0) edge[Green_soc] node[above] {$\omega , \alpha$} (.5,0);
			\end{scope}
		\end{tikzpicture}.
		$\alpha$ indicates the spin direction.
		The direction ensures the right order within the trace (due to the Pauli matrices, the different objects in the diagram do not commute).
		The line corresponds to $\mathcal{M}_{i}(\omega) m^\alpha_i \sigma^\alpha$.
		
	\end{enumerate}
	
	Note that the diagrammatic rules might be counter-intuitive, since local quantities (the Green function) are represented by lines, while non-local quantities (the hopping from $i$ to $j$) are represented by vertices.
	However, these diagrammatic rules allow a much simplified description and identification of all the possible forms of the Gilbert damping, without having to write lengthy perturbative expansions.

	\textbf{Spin-orbit coupling independent contributions.} 
	
	To get a feeling for the diagrammatic approach, we start with the simplest example: the on-site susceptibility without any hoppings to a different site, which describes both the single atom and the lowest order term for interacting atoms.
	The possible forms are,
	\begin{align}
		\chi_{\alpha \beta}^{ii} (\omega) &\propto
		\begin{tikzpicture}[baseline={([yshift=-.5ex]current bounding box.center)}]
			\begin{scope}[scale=2]
				\coordinate (lb) at (0,0);
				\coordinate (mt) at (0.5,-0.5);
				\coordinate (rb) at (1,0);
				\coordinate (lt) at (0,1);
				\coordinate (mt) at (0.5,0.5);
				\coordinate (rt) at (1,1);
				\draw (lb) edge[Green,out=-45,in=-135] node[below,yshift=-.2cm] {$\omega$} (rb);
				\draw (rb) edge[Green,out=135,in=45] node[above,yshift=.2cm] {$0$} (lb);
				\draw[susc] (lb) circle (2pt) node[left] {$i \alpha \ $};
				\draw[susc] (rb) circle (2pt) node[right] {$\ i \beta$};
			\end{scope}
		\end{tikzpicture}
		+
		\begin{tikzpicture}[baseline={([yshift=-.5ex]current bounding box.center)}]
			\begin{scope}[scale=2]
				\coordinate (lb) at (0,0);
				\coordinate (mt) at (0.5,-0.5);
				\coordinate (rb) at (1,0);
				\coordinate (lt) at (0,1);
				\coordinate (mt) at (0.5,0.5);
				\coordinate (rt) at (1,1);
				\draw (lb) edge[Green_soc,out=-45,in=-135] node[below,yshift=-.2cm] {$\omega, \gamma$} (rb);
				\draw (rb) edge[Green,out=135,in=45] node[above,yshift=.2cm] {$0$} (lb);
				\draw[susc] (lb) circle (2pt) node[left] {$i \alpha \ $};
				\draw[susc] (rb) circle (2pt) node[right] {$\ i \beta$};
			\end{scope}
		\end{tikzpicture}\nonumber\\
		&+
		\begin{tikzpicture}[baseline={([yshift=-.5ex]current bounding box.center)}]
			\begin{scope}[scale=2]
				\coordinate (lb) at (0,0);
				\coordinate (mt) at (0.5,-0.5);
				\coordinate (rb) at (1,0);
				\coordinate (lt) at (0,1);
				\coordinate (mt) at (0.5,0.5);
				\coordinate (rt) at (1,1);
				\draw (lb) edge[Green,out=-45,in=-135] node[below,yshift=-.2cm] {$\omega$} (rb);
				\draw (rb) edge[Green_soc,out=135,in=45] node[above,yshift=.2cm] {$0, \gamma$} (lb);
				\draw[susc] (lb) circle (2pt) node[left] {$i \alpha \ $};
				\draw[susc] (rb) circle (2pt) node[right] {$\ i \beta$};
			\end{scope}
		\end{tikzpicture}
		+
		\begin{tikzpicture}[baseline={([yshift=-.5ex]current bounding box.center)}]
			\begin{scope}[scale=2]
				\coordinate (lb) at (0,0);
				\coordinate (mt) at (0.5,-0.5);
				\coordinate (rb) at (1,0);
				\coordinate (lt) at (0,1);
				\coordinate (mt) at (0.5,0.5);
				\coordinate (rt) at (1,1);
				\draw (lb) edge[Green_soc,out=-45,in=-135] node[below,yshift=-.2cm] {$\omega, \delta$} (rb);
				\draw (rb) edge[Green_soc,out=135,in=45] node[above,yshift=.2cm] {$0, \gamma$} (lb);
				\draw[susc] (lb) circle (2pt) node[left] {$i \alpha \ $};
				\draw[susc] (rb) circle (2pt) node[right] {$\ i \beta$};
			\end{scope}
		\end{tikzpicture} \quad ,
	\end{align}
	which evaluate to,
	\begin{align}
		\begin{tikzpicture}[baseline={([yshift=-.5ex]current bounding box.center)}]
			\begin{scope}[scale=2]
				\coordinate (lb) at (0,0);
				\coordinate (mt) at (0.5,-0.5);
				\coordinate (rb) at (1,0);
				\coordinate (lt) at (0,1);
				\coordinate (mt) at (0.5,0.5);
				\coordinate (rt) at (1,1);
				\draw (lb) edge[Green,out=-45,in=-135] node[below,yshift=-.2cm] {$\omega$} (rb);
				\draw (rb) edge[Green,out=135,in=45] node[above,yshift=.2cm] {$0$} (lb);
				\draw[susc] (lb) circle (2pt) node[left] {$i \alpha \ $};
				\draw[susc] (rb) circle (2pt) node[right] {$\ i \beta$};
			\end{scope}
		\end{tikzpicture}
		&= \Tr \sigma^\alpha \sigma^\beta \mathcal{N}_{i}\omega) \mathcal{N}_{i}(0) = \delta_{\alpha \beta} \mathcal{N}_{i}(\omega) \mathcal{N}_{i}(0)
		\\
		\begin{tikzpicture}[baseline={([yshift=-.5ex]current bounding box.center)}]
			\begin{scope}[scale=2]
				\coordinate (lb) at (0,0);
				\coordinate (mt) at (0.5,-0.5);
				\coordinate (rb) at (1,0);
				\coordinate (lt) at (0,1);
				\coordinate (mt) at (0.5,0.5);
				\coordinate (rt) at (1,1);
				\draw (lb) edge[Green_soc,out=-45,in=-135] node[below,yshift=-.2cm] {$\omega, \gamma$} (rb);
				\draw (rb) edge[Green,out=135,in=45] node[above,yshift=.2cm] {$0$} (lb);
				\draw[susc] (lb) circle (2pt) node[left] {$i \alpha \ $};
				\draw[susc] (rb) circle (2pt) node[right] {$\ i \beta$};
			\end{scope}
		\end{tikzpicture}
		&=
		\Tr \sigma^\alpha \sigma^\gamma \sigma^\beta \mathcal{M}_{i}(\omega) \mathcal{N}_{i}(0) m_i^\gamma= \iu \epsilon_{\alpha \gamma \beta} \mathcal{M}_{i}(\omega) \mathcal{N}_{i}(0) m_i^\gamma \label{eq:diagramm_anti_sym_1}
		\\
		\begin{tikzpicture}[baseline={([yshift=-.5ex]current bounding box.center)}]
			\begin{scope}[scale=2]
				\draw (lb) edge[Green,out=-45,in=-135] node[below,yshift=-.2cm] {$\omega$} (rb);
				\draw (rb) edge[Green_soc,out=135,in=45] node[above,yshift=.2cm] {$0, \gamma$} (lb);
				\draw[susc] (lb) circle (2pt) node[left] {$i \alpha \ $};
				\draw[susc] (rb) circle (2pt) node[right] {$\ i \beta$};
			\end{scope}
		\end{tikzpicture}
		&=
		\Tr \sigma^\alpha \sigma^\beta \sigma^\gamma \mathcal{N}_{i}(\omega) \mathcal{M}_{i}(0) m_i^\gamma= \iu \epsilon_{\alpha \beta \gamma} \mathcal{M}_{i}(\omega) \mathcal{M}_{i}(0) m_i^\gamma \label{eq:diagramm_anti_sym_2}
		\\
		\begin{tikzpicture}[baseline={([yshift=-.5ex]current bounding box.center)}]
			\begin{scope}[scale=2]
				\draw (lb) edge[Green_soc,out=-45,in=-135] node[below,yshift=-.2cm] {$\omega, \delta$} (rb);
				\draw (rb) edge[Green_soc,out=135,in=45] node[above,yshift=.2cm] {$0, \gamma$} (lb);
				\draw[susc] (lb) circle (2pt) node[left] {$i \alpha \ $};
				\draw[susc] (rb) circle (2pt) node[right] {$\ i \beta$};
			\end{scope}
		\end{tikzpicture}
		&=
		\Tr \sigma^\alpha \sigma^\delta \sigma^\beta \sigma^\gamma \mathcal{M}_{i}(\omega) \mathcal{M}_{i}(0) m_i^\delta m_i^\gamma
		\nonumber\\
		&=\left( \delta_{\alpha \delta} \delta_{\beta \gamma} + \delta_{\alpha \gamma} \delta_{\beta \delta} - \delta_{\alpha\beta} \delta_{\gamma \delta}  \right) \mathcal{M}_{i}(\omega) \mathcal{M}_{i}(0) m_i^\delta m_i^\gamma \quad .
	\end{align}
	The first diagram yields an isotropic contribution, the second and third diagrams yield an anti-symmetric contribution, which is linear in the magnetic moment, and the last diagram yields a symmetric contribution being quadratic in the magnetic moment.
	Note that the energy dependence of the Green functions is crucial, since otherwise the sum of eqs. \eqref{eq:diagramm_anti_sym_1} and \eqref{eq:diagramm_anti_sym_2} vanishes.
	In particular this means that the static susceptibility has no dependence linear in the magnetic moment, while the the slope of the susceptibility with respect to energy can have a dependence linear in the magnetic moment.
	The static part of the susceptibility maps to the magnetic exchange interactions, which are known to be even in the magnetic moment due to time reversal symmetry. 
	
	Combining all the functional forms of the diagrams, we find the following possible dependencies of the on-site Gilbert damping on the magnetic moments,
	\begin{align}
		\mathcal{G}_{ii}^{\alpha \beta}(\{ \vec{m} \}) \propto \{ \delta_{\alpha \beta}, \epsilon_{\alpha \beta \gamma} m_i^\gamma , m_i^\alpha m_i^\beta\} \quad . \label{eq:damping_on-site_single}
	\end{align}
	Since we work in the local frames, $\vec{m}_i = ( 0, 0, m_i^z )$, the last dependence is a purely longitudinal term, which is not relevant for the transversal dynamics discussed in this work.
	
	If we still focus on the on-site term, but allow for two hoppings to another atom and back, we find the following new diagrams,
	\begin{align}
		&\begin{tikzpicture}[baseline={([yshift=-.5ex]current bounding box.center)}]
			\begin{scope}[scale=2]
				\draw (lb) edge[Green] node[below,yshift=-.2cm] {$\omega$} (rb);
				\draw (rb) edge[Green] node[right,xshift=.2cm] {$0$} (rt);
				\draw (rt) edge[Green] node[above,yshift=.2cm] {$0$} (lt);
				\draw (lt) edge[Green] node[left,yshift=-.2cm] {$0$} (lb);
				\draw[susc] (lb) circle (2pt) node[left] {$i \alpha \ $};
				\draw[susc] (rb) circle (2pt) node[right] {$\ i \beta$};
				\draw[dt] (rt) circle (2pt) node[right] {$\ ij$};
				\draw[dt] (lt) circle (2pt) node[left] {$ ji \ $};
			\end{scope}
		\end{tikzpicture} 
		+
		\begin{tikzpicture}[baseline={([yshift=-.5ex]current bounding box.center)}]
			\begin{scope}[scale=2]
				\draw (lb) edge[Green_soc] node[below,yshift=-.2cm] {$\omega, \gamma$} (rb);
				\draw (rb) edge[Green] node[right,xshift=.2cm] {$0$} (rt);
				\draw (rt) edge[Green] node[above,yshift=.2cm] {$0$} (lt);
				\draw (lt) edge[Green] node[left,yshift=-.2cm] {$0$} (lb);
				\draw[susc] (lb) circle (2pt) node[left] {$i \alpha \ $};
				\draw[susc] (rb) circle (2pt) node[right] {$\ i \beta$};
				\draw[dt] (rt) circle (2pt) node[right] {$\ ij$};
				\draw[dt] (lt) circle (2pt) node[left] {$ ji \ $};
			\end{scope}
		\end{tikzpicture}
		+ \hdots +
		\begin{tikzpicture}[baseline={([yshift=-.5ex]current bounding box.center)}]
			\begin{scope}[scale=2]
				\draw (lb) edge[Green_soc] node[below,yshift=-.2cm] {$\omega, \gamma$} (rb);
				\draw (rb) edge[Green_soc] node[right,xshift=.2cm] {$0, \delta$} (rt);
				\draw (rt) edge[Green] node[above,yshift=.2cm] {$0$} (lt);
				\draw (lt) edge[Green] node[left,yshift=-.2cm] {$0$} (lb);
				\draw[susc] (lb) circle (2pt) node[left] {$i \alpha \ $};
				\draw[susc] (rb) circle (2pt) node[right] {$\ i \beta$};
				\draw[dt] (rt) circle (2pt) node[right] {$\ ij$};
				\draw[dt] (lt) circle (2pt) node[left] {$ ji \ $};
			\end{scope}
		\end{tikzpicture}
		+ \hdots \nonumber \\
		&+
		\begin{tikzpicture}[baseline={([yshift=-.5ex]current bounding box.center)}]
			\begin{scope}[scale=2]
				\draw (lb) edge[Green_soc] node[below,yshift=-.2cm] {$\omega, \gamma$} (rb);
				\draw (rb) edge[Green_soc] node[right,xshift=.2cm] {$0, \delta$} (rt);
				\draw (rt) edge[Green_soc] node[above,yshift=.2cm] {$0, \eta$} (lt);
				\draw (lt) edge[Green] node[left,yshift=-.2cm] {$0$} (lb);
				\draw[susc] (lb) circle (2pt) node[left] {$i \alpha \ $};
				\draw[susc] (rb) circle (2pt) node[right] {$\ i \beta$};
				\draw[dt] (rt) circle (2pt) node[right] {$\ ij$};
				\draw[dt] (lt) circle (2pt) node[left] {$ ji \ $};
			\end{scope}
		\end{tikzpicture}
		+ \hdots +
		\begin{tikzpicture}[baseline={([yshift=-.5ex]current bounding box.center)}]
			\begin{scope}[scale=2]
				\draw (lb) edge[Green_soc] node[below,yshift=-.2cm] {$\omega, \gamma$} (rb);
				\draw (rb) edge[Green_soc] node[right,xshift=.2cm] {$0, \delta$} (rt);
				\draw (rt) edge[Green_soc] node[above,yshift=.2cm] {$0, \eta$} (lt);
				\draw (lt) edge[Green_soc] node[left,yshift=-.2cm] {$0, \nu$} (lb);
				\draw[susc] (lb) circle (2pt) node[left] {$i \alpha \ $};
				\draw[susc] (rb) circle (2pt) node[right] {$\ i \beta$};
				\draw[dt] (rt) circle (2pt) node[right] {$\ ij$};
				\draw[dt] (lt) circle (2pt) node[left] {$ ji \ $};
			\end{scope}
		\end{tikzpicture} \quad .
	\end{align}
	The dashed line in the second diagram can be inserted in any of the four sides of the square, with the other possibilities omitted.
	Likewise for the diagrams with two or three dashed lines, the different possible assignments have to be considered.
	The additional hopping to the site $j$ yields a dependence of the on-site magnetic susceptibility and therefore also the on-site Gilbert damping tensor on the magnetic moment of site $j$.
	
	Another contribution to the Gilbert damping originates from the inter-site part, thus encoding the dependence of the moment site $i$ on the dynamics of the moment of site $j$ via $\mathcal{G}_{ij}$.
	This contribution is often neglected in the  literature, since for many systems it is believed to have no significant impact.
	Using the microscopic model, a different class of diagrams is responsible for the inter-site damping.
	In the lowest order in $t/Um$ the diagrams contain already two hopping events,
	\begin{align}
		&\begin{tikzpicture}[baseline={([yshift=-.5ex]current bounding box.center)}]
			\begin{scope}[scale=2]
				\coordinate (lb) at (0,0);
				\coordinate (mb) at (0.5,-0.5);
				\coordinate (rb) at (1,0);
				\coordinate (lt) at (0,1);
				\coordinate (mt) at (0.5,0.5);
				\coordinate (rt) at (1,1);
				\draw (lb) edge[Green] node[left,xshift=-.2cm] {$\omega$} (mb);
				\draw (mb) edge[Green] node[right,xshift=.2cm] {$\omega$} (rb);
				\draw (rb) edge[Green] node[right,yshift=.2cm] {$0$} (mt);
				\draw (mt) edge[Green] node[left,yshift=.2cm] {$0$} (lb);
				\draw[susc] (lb) circle (2pt) node[left] {$i \alpha \ $};
				\draw[susc] (rb) circle (2pt) node[right] {$\ j \beta$};
				\draw[dt] (mb) circle (2pt) node[below,yshift=-.2cm] {$ij$};
				\draw[dt] (mt) circle (2pt) node[above,yshift=.2cm] {$ij$};
			\end{scope}
		\end{tikzpicture}
		+ 
		\begin{tikzpicture}[baseline={([yshift=-.5ex]current bounding box.center)}]
			\begin{scope}[scale=2]
				\coordinate (lb) at (0,0);
				\coordinate (mb) at (0.5,-0.5);
				\coordinate (rb) at (1,0);
				\coordinate (lt) at (0,1);
				\coordinate (mt) at (0.5,0.5);
				\coordinate (rt) at (1,1);
				\draw (lb) edge[Green_soc] node[left,xshift=-.2cm] {$\omega, \gamma$} (mb);
				\draw (mb) edge[Green] node[right,xshift=.2cm] {$\omega$} (rb);
				\draw (rb) edge[Green] node[right,yshift=.2cm] {$0$} (mt);
				\draw (mt) edge[Green] node[left,yshift=.2cm] {$0$} (lb);
				\draw[susc] (lb) circle (2pt) node[left] {$i \alpha \ $};
				\draw[susc] (rb) circle (2pt) node[right] {$\ j \beta$};
				\draw[dt] (mb) circle (2pt) node[below,yshift=-.2cm] {$ij$};
				\draw[dt] (mt) circle (2pt) node[above,yshift=.2cm] {$ij$};
			\end{scope}
		\end{tikzpicture}
		+ \hdots 
		+
		\begin{tikzpicture}[baseline={([yshift=-.5ex]current bounding box.center)}]
			\begin{scope}[scale=2]
				\coordinate (lb) at (0,0);
				\coordinate (mb) at (0.5,-0.5);
				\coordinate (rb) at (1,0);
				\coordinate (lt) at (0,1);
				\coordinate (mt) at (0.5,0.5);
				\coordinate (rt) at (1,1);
				\draw (lb) edge[Green_soc] node[left,xshift=-.2cm] {$\omega,\gamma$} (mb);
				\draw (mb) edge[Green_soc] node[right,xshift=.2cm] {$\omega,\delta$} (rb);
				\draw (rb) edge[Green] node[right,yshift=.2cm] {$0$} (mt);
				\draw (mt) edge[Green] node[left,yshift=.2cm] {$0$} (lb);
				\draw[susc] (lb) circle (2pt) node[left] {$i \alpha \ $};
				\draw[susc] (rb) circle (2pt) node[right] {$\ j \beta$};
				\draw[dt] (mb) circle (2pt) node[below,yshift=-.2cm] {$ij$};
				\draw[dt] (mt) circle (2pt) node[above,yshift=.2cm] {$ij$};
			\end{scope}
		\end{tikzpicture}
		+ \hdots \nonumber \\
		&+
		\begin{tikzpicture}[baseline={([yshift=-.5ex]current bounding box.center)}]
			\begin{scope}[scale=2]
				\coordinate (lb) at (0,0);
				\coordinate (mb) at (0.5,-0.5);
				\coordinate (rb) at (1,0);
				\coordinate (lt) at (0,1);
				\coordinate (mt) at (0.5,0.5);
				\coordinate (rt) at (1,1);
				\draw (lb) edge[Green_soc] node[left,xshift=-.2cm] {$\omega,\gamma$} (mb);
				\draw (mb) edge[Green_soc] node[right,xshift=.2cm] {$\omega,\delta$} (rb);
				\draw (rb) edge[Green_soc] node[right,yshift=.2cm] {$0,\eta$} (mt);
				\draw (mt) edge[Green] node[left,yshift=.2cm] {$0$} (lb);
				\draw[susc] (lb) circle (2pt) node[left] {$i \alpha \ $};
				\draw[susc] (rb) circle (2pt) node[right] {$\ j \beta$};
				\draw[dt] (mb) circle (2pt) node[below,yshift=-.2cm] {$ij$};
				\draw[dt] (mt) circle (2pt) node[above,yshift=.2cm] {$ij$};
			\end{scope}
		\end{tikzpicture}
		+ \hdots 
		+
		\begin{tikzpicture}[baseline={([yshift=-.5ex]current bounding box.center)}]
			\begin{scope}[scale=2]
				\coordinate (lb) at (0,0);
				\coordinate (mb) at (0.5,-0.5);
				\coordinate (rb) at (1,0);
				\coordinate (lt) at (0,1);
				\coordinate (mt) at (0.5,0.5);
				\coordinate (rt) at (1,1);
				\draw (lb) edge[Green_soc] node[left,xshift=-.2cm] {$\omega,\gamma$} (mb);
				\draw (mb) edge[Green_soc] node[right,xshift=.2cm] {$\omega,\delta$} (rb);
				\draw (rb) edge[Green_soc] node[right,yshift=.2cm] {$0,\eta$} (mt);
				\draw (mt) edge[Green_soc] node[left,yshift=.2cm] {$0,\zeta$} (lb);
				\draw[susc] (lb) circle (2pt) node[left] {$i \alpha \ $};
				\draw[susc] (rb) circle (2pt) node[right] {$\ j \beta$};
				\draw[dt] (mb) circle (2pt) node[below,yshift=-.2cm] {$ij$};
				\draw[dt] (mt) circle (2pt) node[above,yshift=.2cm] {$ij$};
			\end{scope}
		\end{tikzpicture} \quad . \label{eq:intersite_diagrams_second_order}
	\end{align}

	In total, we find that the spin-orbit independent intra-site and inter-site Gilbert damping tensors can be respectively written as 
	\begin{align}
		\begin{split}
			\mathcal{G}_{ii} =& 
			\left( \mathcal{S}_i + \mathcal{S}_i^{ij,(1)} \left(\vec{e}_i \cdot \vec{e}_j \right) + \mathcal{S}_i^{ij,(2)} \left(\vec{e}_i \cdot \vec{e}_j \right)^2 \right) \mathcal{I} \\+& \left( \mathcal{A}_i + \mathcal{A}_i^{ij}  \left(\vec{e}_i \cdot \vec{e}_j \right) \right) \mathcal{E}(\vec{e}_i),
		\end{split}
	\end{align}
	and
	\begin{align}
		\begin{split}
			\mathcal{G}_{ij}^{\alpha\beta} =& 
			\left( \mathcal{S}_{ij} + \mathcal{S}_{ij}^{\mathrm{dot}} \left(\vec{e}_i \cdot \vec{e}_j \right)  \right)\delta_{\alpha\beta} \\+& \left( \mathcal{A}_{ij} + \mathcal{A}_{ij}^{\mathrm{dot}}  \left(\vec{e}_i \cdot \vec{e}_j \right) \right) \left( \mathcal{E}(\vec{e}_i) + \mathcal{E}(\vec{e}_j) \right)^{\alpha\beta}\\ +& \mathcal{S}_{ij}^{\mathrm{cross}} \left( \vec{e}_i \times \vec{e}_j \right)^\alpha \left( \vec{e}_i \times \vec{e}_j \right)^\beta 
			+ \mathcal{S}_{ij}^{\mathrm{ba}} e_i^\beta e_j^\alpha
		\end{split} \quad, \label{eq:inter-site_damping}
	\end{align}
	where as mentioned earlier $\mathcal{S}$ and $\mathcal{A}$ represent symmetric and asymmetric contributions, $\mathcal{I}$ is the $3\times3$ identity while $\mathcal{E}(\vec{e}_i) = \begin{pmatrix} 0 & e_i^z & -e_i^y \\ -e_i^z & 0 & e_i^x \\ e_i^y & -e_i^x & 0 \end{pmatrix}$.  
	
	Remarkably, we find that both the symmetric and anti-symmetric parts of the Gilbert damping tensor have a rich dependence with the opening angle of the magnetic moments. We identify, for example, the dot and the square of the dot products of the magnetic moments to possibly play a crucial role in modifying the damping, similarly to bilinear and biquadratic magnetic interactions. It is worth noting that even though the intra-site Gilbert damping can explicitly depend on other magnetic moments, its meaning remains unchanged.
	The anti-symmetric precession-like term describes a precession of the moment around its own effective magnetic field, while the diagonal damping-like term describes a damping towards its own effective magnetic field.
	The dependence on other magnetic moments renormalizes the intensity of those two processes.
	The inter-site Gilbert damping describes similar processes, but with respect to the effective field of the other involved magnetic moment. On the basis of the LLG equation, eq.~\eqref{eq:general_LLG}, it can be shown that the term related to $\mathcal{S}_{ij}^\text{ba}$ with a functional form of $ e_i^\beta e_j^\alpha$ describes a precession of the $i$-th moment around the $j$-th moment with a time- and directional-dependent amplitude, $ \partial_t \vec{m}_i \propto \left( \vec{m}_i \times  \vec{m}_j \right) \left(  \vec{m}_i \cdot \partial_t  \vec{m}_j \right) $.
	The double cross product term yields a time dependence of $ \partial_t \vec{m}_i \propto \left( \vec{m}_i \times \left( \vec{m}_i \times  \vec{m}_j \right) \right) \left(  \left( \vec{m}_i \times  \vec{m}_j \right) \cdot \partial_t  \vec{m}_j \right) $.
	Both contributions are neither pure precession-like nor pure damping-like, but show complex time- and directional-dependent dynamics.
	
	\textbf{Spin-orbit coupling contributions.}
	The spin-orbit interaction gives rise to new possible dependencies of the damping on the magnetic structure.
	In particular, the so-called chiral damping, which in general is the difference of the damping between a right-handed and a left-handed opening, rises from SOC and broken inversion symmetry.
	Using our perturbative model, we can identify all possible dependencies up to second order in SOC and third order in the magnetic moments. 
	
	In the diagramms SOC is added by replacing one spin-independent hopping vertex by a spin-dependent one,
	\begin{align}
		\begin{tikzpicture}[baseline={([yshift=-.5ex]current bounding box.center)}]
			\begin{scope}[scale=2]
				\draw (lb) edge[Green] node[below,yshift=-.2cm] {$\omega$} (rb);
				\draw (rb) edge[Green] node[right,xshift=.2cm] {$0$} (rt);
				\draw (rt) edge[Green] node[above,yshift=.2cm] {$0$} (lt);
				\draw (lt) edge[Green] node[left,yshift=-.2cm] {$0$} (lb);
				\draw[susc] (lb) circle (2pt) node[left] {$i \alpha \ $};
				\draw[susc] (rb) circle (2pt) node[right] {$\ i \beta$};
				\draw[dt] (rt) circle (2pt) node[right] {$\ ij$};
				\draw[dt] (lt) circle (2pt) node[left] {$ ji \ $};
			\end{scope}
		\end{tikzpicture} 
		\rightarrow
		\begin{tikzpicture}[baseline={([yshift=-.5ex]current bounding box.center)}]
			\begin{scope}[scale=2]
				\draw (lb) edge[Green] node[below,yshift=-.2cm] {$\omega$} (rb);
				\draw (rb) edge[Green] node[right,xshift=.2cm] {$0$} (rt);
				\draw (rt) edge[Green] node[above,yshift=.2cm] {$0$} (lt);
				\draw (lt) edge[Green] node[left,yshift=-.2cm] {$0$} (lb);
				\draw[susc] (lb) circle (2pt) node[left] {$i \alpha \ $};
				\draw[susc] (rb) circle (2pt) node[right] {$\ i \beta$};
				\draw[soc] (rt) circle (2pt) node[right] {$\ ij \gamma$};
				\draw[dt] (lt) circle (2pt) node[left] {$ ij \ $};
			\end{scope}
		\end{tikzpicture} \quad . \label{eq:soc_non_coll_susc_diagramm}
	\end{align}
	
	Up to first-order in SOC, we find the  the following dependencies were found for the on-site Gilbert damping
	\begin{align}
		\mathcal{G}_{ii}(\{\vec{m}\}) \propto \{&
		\epsilon_{\alpha \beta \gamma} \hat{n}_{ij}^\gamma ,
		\hat{n}_{ij}^\alpha \hat{n}_{ji}^\beta ,
		\hat{n}_{ij}^\beta m_i^\alpha ,
		\hat{n}_{ij}^\alpha m_i^\beta ,
		\delta_{\alpha \beta} (\hat{\vec{n}}_{ij} \cdot \vec{m}_i) ,
		\delta_{\alpha \beta} (\hat{\vec{n}}_{ij} \cdot \vec{m}_j) ,
		\nonumber \\
		&\hat{n}_{ij}^\beta m_j^\alpha ,
		\hat{n}_{ij}^\alpha m_j^\beta ,
		m_i^\alpha (\hat{\vec{n}}_{ij} \times \vec{m}_i)^\beta ,
		m_i^\beta (\hat{\vec{n}}_{ij} \times \vec{m}_i)^\alpha ,
		\nonumber \\
		&\delta_{\alpha\beta} \hat{\vec{n}}_{ij} \cdot (\vec{m}_i \times \vec{m}_j),
		m_i^\alpha (\hat{\vec{n}}_{ij} \times \vec{m}_j)^\beta ,
		m_i^\beta (\hat{\vec{n}}_{ij} \times \vec{m}_j)^\alpha ,
		(\hat{\vec{n}}_{ij} \cdot \vec{m}_j) \epsilon_{\alpha \beta \gamma} m_i^\gamma ,
		\nonumber \\
		&m_i^\alpha m_i^\beta (\hat{\vec{n}}_{ij} \cdot \vec{m}_j) ,
		(m_i^\alpha m_j^\beta - m_i^\beta m_j^\alpha) (\hat{\vec{n}}_{ij} \cdot \vec{m}_j ) ,
		\hat{n}_{ij}^\beta m_i^\alpha (\vec{m}_i \cdot \vec{m}_j) ,
		\hat{n}_{ij}^\alpha m_i^\beta (\vec{m}_i \cdot \vec{m}_j) \} \quad . \label{eq:second_order_dependencies_on_site_soc}
	\end{align}

	We identified the following contributions for the on-site and intersite damping to be the most relevant one after the numerical evaluation discussed in the next sections:
	\begin{align}
		\mathcal{G}_{ii}^{\text{soc}} =& 
		\mathcal{S}_i^{\text{soc},ij} \ \vec{n}_{ij}\cdot ( \vec{e}_i \times \vec{e}_j ) \mathcal{I} \nonumber\\+&
		\mathcal{S}_i^{\text{soc},ij,(2)} \ (\vec{n}_{ij}\cdot  \vec{e}_i)(\vec{n}_{ij}\cdot  \vec{e}_j) \mathcal{I} \nonumber\\+ & \mathcal{A}_i^{\text{soc},ij} \ \vec{n}_{ij}\cdot ( \vec{e}_i \times \vec{e}_j ) \mathcal{E}(\vec{e}_i) \nonumber\\+& \mathcal{A}_i^{\text{soc},ij,(2)} \ (\vec{n}_{ij}\cdot \vec{e}_j ) \mathcal{E}(\vec{n}_{ij}), 
	\end{align}
	and
	\begin{align}
		\mathcal{G}_{ij}^{\text{soc},\alpha\beta} =& \mathcal{S}_{ij}^{\text{soc}} \ \vec{n}_{ij}\cdot ( \vec{e}_i \times \vec{e}_j ) \delta_{\alpha\beta} +
		\mathcal{S}_{ij}^{\text{soc},ba} n_{ij}^\beta ( \vec{e}_i \times \vec{e}_j )^\alpha \nonumber \\ +& \mathcal{A}_{ij}^{\text{soc}} \mathcal{E}^{\alpha\beta}(\vec{n}_{ij}) \quad . \label{eq:inter-site_damping_soc}
	\end{align}
	The contributions being first-order in SOC are obviously chiral since they depend on the cross product, $\vec{e}_i \times \vec{e}_j$. Thus, similar to the magnetic Dzyaloshinskii-Moriya interaction, SOC gives rise to a dependence of the Gilbert damping on the vector chirality, $\vec{e}_i \times \vec{e}_j$.
	The term chiral damping used in literature refers to the dependence of the Gilbert damping on the chirality, but to our knowledge it was not shown so far how this dependence evolves from a microscopic model, and how it looks like in an atomistic model.

	\textbf{Extension to three sites.}
	Including three different sites $i$, $j$, and $k$ in the expansions allows for a ring exchange $i\rightarrow j \rightarrow k \rightarrow i$ involving three hopping processes, which gives rise to new dependencies of the Gilbert damping on the directions of the moments.
	
	An example of a diagram showing up for the on-site Gilbert damping is given below for the on-site Gilbert damping the diagram,
	\begin{align}
		\begin{tikzpicture}[baseline={([yshift=-.5ex]current bounding box.center)}]
			\begin{scope}[scale=2]
				\coordinate (lb) at (0,0);
				\coordinate (mb) at (0.5,-0.5);
				\coordinate (rb) at (1,0);
				\coordinate (lt) at (0,.75);
				\coordinate (mt) at (0.5,1.25);
				\coordinate (rt) at (1,.75);
				\draw (lb) edge[Green] node[below,yshift=-.2cm] {$\omega$} (rb);
				\draw (rb) edge[Green] node[right,xshift=.2cm] {$0$} (rt);
				\draw (rt) edge[Green] node[right,xshift=.2cm] {$0$} (mt);
				\draw (mt) edge[Green] node[left,xshift=-.2cm] {$0,\gamma$} (lt);
				\draw (lt) edge[Green] node[left,xshift=-.2cm] {$0$} (lb);
				\draw[susc] (lb) circle (2pt) node[left] {$i \alpha \ $};
				\draw[susc] (rb) circle (2pt) node[right] {$\ i \beta$};
				\draw[dt] (rt) circle (2pt) node[right,xshift=.2cm] {$ij$};
				\draw[dt] (mt) circle (2pt) node[above,yshift=.2cm] {$jk$};
				\draw[dt] (lt) circle (2pt) node[left,xshift=-.2cm] {$ki$};
			\end{scope}
		\end{tikzpicture}
	\end{align}
	
	Apart from the natural extensions of the previously discussed 2-site quantities, the intra-site Gilbert damping of site $i$ can depend on the angle between the sites $j$ and $k$, $\vec{e}_j \cdot \vec{e}_k$, or in higher-order on the product of the angles between site $i$ and $j$ with $i$ and $k$, $(\vec{e}_i \cdot \vec{e}_j) (\vec{e}_i \cdot \vec{e}_k)$.
	In sixth-order in the magnetic moments the term $(\vec{e}_i \cdot \vec{e}_j) (\vec{e}_j \cdot \vec{e}_k) (\vec{e}_k \cdot \vec{e}_i)$ yields to a dependence on the square of the scalar spin chirality of the three sites, $\left[\vec{e}_i \cdot (\vec{e}_j \times \vec{e}_k )\right]^2$. 
	Including SOC, there are two interesting dependencies on the scalar spin chirality.
	In first-order one finds similarly to the recently discovered chiral biquadratic interaction~\cite{Brinker2019} and its 3-site generalization~\cite{Laszloffy2019}, e.g.~$\left(\vec{n}_{ij}\cdot \vec{e}_i\right) \left(\vec{e}_i \cdot (\vec{e}_j \times \vec{e}_k )\right)$, while in second order a direct dependence on the scalar spin chirality is allowed, e.g.~$n_{ij}^\alpha n_{ki}^\beta \left(\vec{e}_i \cdot (\vec{e}_j \times \vec{e}_k) \right)$. 
	The scalar spin chirality directly relates to the topological orbital moment~\cite{Dias2016,Hanke2016,Dias2017} and therefore the physical origin of those dependencies lies in the topological orbital moment.
	Even though these terms might not be the most important ones in our model, for specific non-collinear configurations or for some realistic elements with a large topological orbital moment, e.g.~MnGe~\cite{Grytsiuk2020}, they might be important and even dominant yielding interesting new physics.

	\section{Application to the Alexander-Anderson model} \label{sec:Anderson_model_results}
	\textbf{Magnetic dimers.} 
	Based on a 2-site Alexander-Anderson model, we investigated the dependence of the Gilbert damping on the directions of the magnetic moments using the previously discussed possible terms (see more details on the method in Appendix~\ref{app:Method_Alexander}).
	The spin splitting $U$ defines the energy scale and all other parameters. The energy of orbitals is set to $E_d = 1.0$. 
	The magnetization is self-consistently constrained  in a range of $m=\SI{0.2}{}$ to $m=\SI{0.8}{}$ using magnetic constraining fields. The corresponding spin-resolved local density of states is illustrated in Fig.~1b, where  the inter-site hopping is set to $t=\SI{0.2}{}$ and the hybridization to $\Gamma = \SI{0.2}{}$.
	We performed two sets of calculations: one without spin-dependent hopping, $\varphi_\text{R}=\SI{0}{\degree}$, and one with a spin-dependent hopping, $\varphi_\text{R}=\SI{20}{\degree}$.
	
	\begin{figure*}[!tb]
		\begin{center}
			\includegraphics[width=\textwidth]{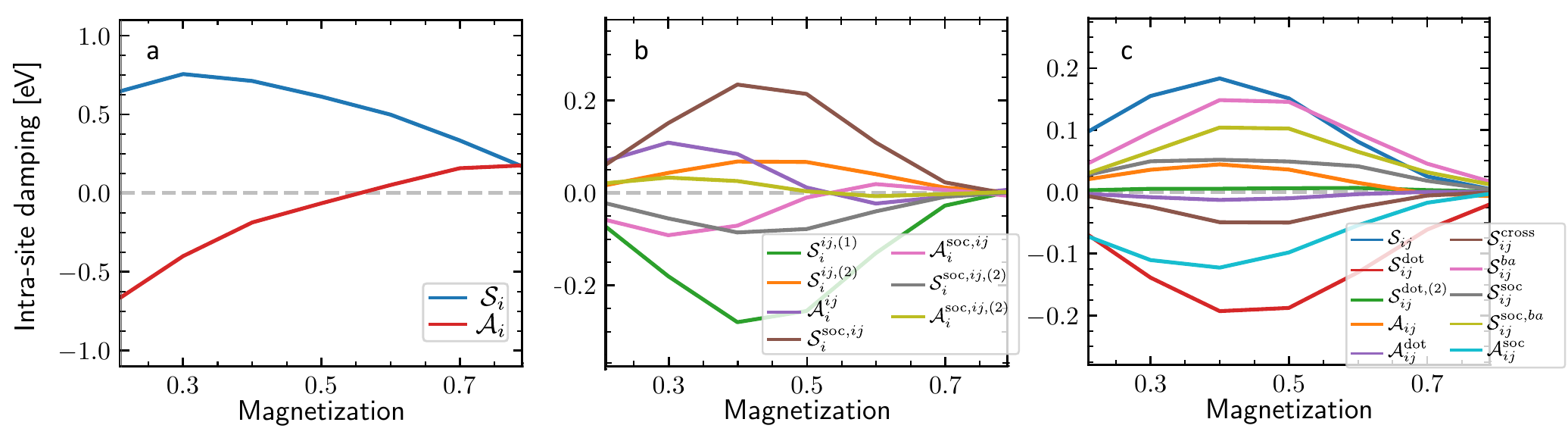}
			\caption{\label{fig:figure2}
				Damping parameters as function of the magnetization for the dimers described within the Alexander-Anderson model including spin orbit coupling.
				A longitudinal magnetic field is used to self-consistently constrain the magnetization. 
				The parameters are extracted from fitting to the inverse of the transversal susceptibility for several non-collinear configurations based on a Lebedev mesh.
				Model parameters in units of $U$: $E_d = \SI{1.0}{}, \ t = \SI{0.2}{}, \ \Gamma = \SI{0.2}{} \ , \ \varphi_\mathrm{R}=\SI{20}{\degree}$.
			}
		\end{center}
	\end{figure*}
	
	The different damping parameters are shown in Fig.~\ref{fig:figure2} as function of the magnetization.
	They are obtained from a least-squares fit to several non-collinear configurations based on a Lebedev mesh for $\ell=2$~\cite{Lebedev1999}. 
	The damping, which is independent of the relative orientation of the two sites, is shown in Fig.~\ref{fig:figure2}a.
	The symmetric damping-like intra-site contribution $\mathcal{S}_i$ dominates the damping tensor for most magnetizations and has a maximum at $m = \SI{0.3}{}$.
	The anti-symmetric intra-site contribtuion $\mathcal{A}_i$, which renormalizes the gyromagnetic ratio, approximately changes sign when the Fermi level passes the peak of the minority spin channel at $m \approx \SI{0.5}{}$ and has a significantly larger amplitude for small magnetizations.
	Both contributions depend mainly on the broadening $\Gamma$, which mimics the coupling to an electron bath and is responsible for the absorption of spin currents, which in turn are responsible for the damping of the magnetization dynamics~\cite{Mizukami2002,Tserkovnyak2002}. 
	
	The directional dependencies of the intra-site damping are shown in Fig.~\ref{fig:figure2}b.
	With our choice of parameters, the correction to the damping-like symmetric Gilbert damping can reach half of the direction-independent term.
	This means that the damping can vary between $\approx 0.4 - 1.0$ for a ferromagnetic and an antiferromagnetic state at $m=\SI{0.4}{}$.
	Also for the renormalization of the gyromagnetic ratio a significant correction is found, which in the ferromagnetic case always lowers and in the antiferromagnetic case enhances the amplitude. 
	The most dominant contribution induced by SOC is the chiral one, which depends on the cross product of the moments $i$ and $j$,   which in terms of amplitude is comparable to the isotropic dot product terms.
	Interestingly, while the inter-site damping term  is in general known to be less relevant than the intra-site damping, we find that this does not hold for the directional dependence of the damping.
	The inter-site damping is shown in Fig.~\ref{fig:figure2}c.
	Even though the directional-independent term, $\mathcal{S}_{ij}$, is nearly one order of magnitude smaller than the equivalent intra-site contribution, this is not necessarily the case  for the directional-dependent terms, which are comparable to the intra-site equivalents.

	\section{Application to first-principles simulations}
	
\begin{figure*}[!tb]
		\begin{center}
			\includegraphics[scale=.9
			]{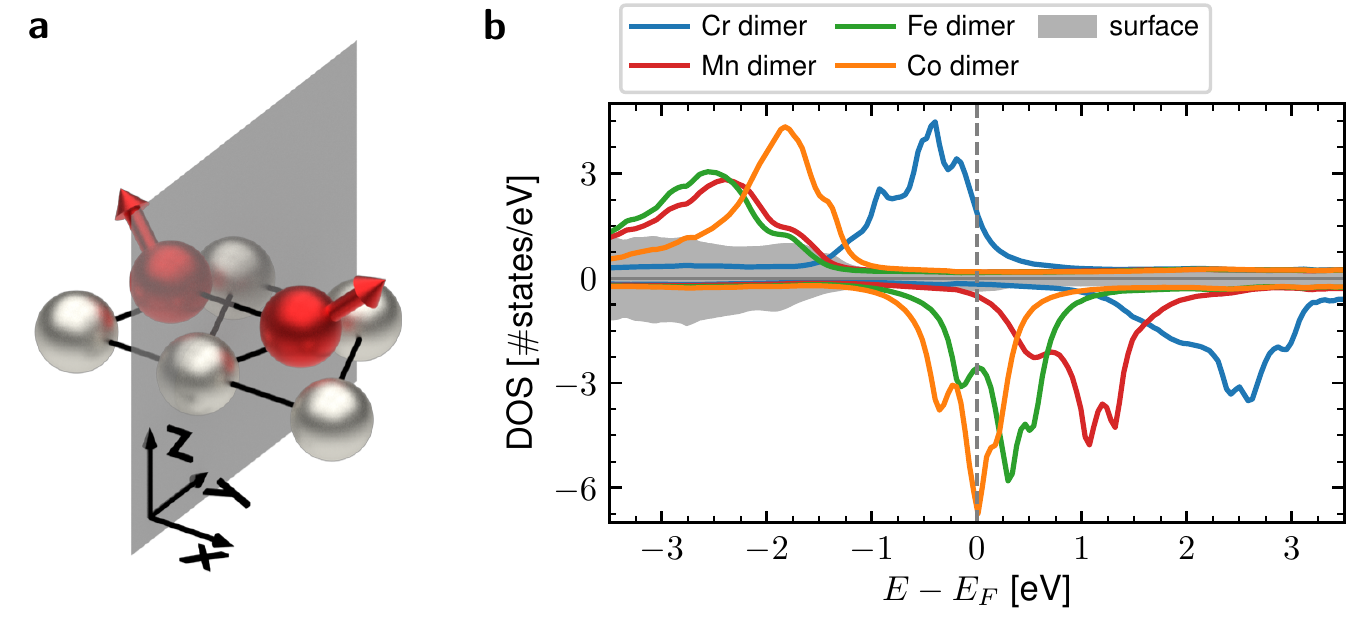}
			\caption{\label{fig:DOS_Au111_dimers} 	
				\textbf{a} Illustration of a non-collinear magnetic dimer (red spheres) deposited on the (111) facets of Au (grey spheres).
				From the initial $C_{3v}$ spatial symmetry of the surface the dimers preserve the mirror plane (indicated grey) in the $y$-$z$ plane.
				\textbf{b} 
				Local density of states of the Cr, Mn, Fe and Co dimers deposited on the Au(111) surface.
				The grey background indicates the surface density of states.
				The dimers are collinear in the $z$-direction.
				\label{fig:figure3}	}
		\end{center}
	\end{figure*}

	To investigate the importance of non-collinear effects for the Gilbert damping in realistic systems, we use DFT and time-dependent DFT to explore the prototypical example of monoatomic $3d$ transition metal adatoms and dimers deposited on a heavy metal surface hosting large SOC (see Fig.~\ref{fig:DOS_Au111_dimers}a for an illustration of the configuration).
	We consider a Cr, Mn, Fe and Co atoms deposited on the fcc-Au(111) surface (details of the simulations are described in Appendix~\ref{app:method_DFT}). 
	The parameters and the corresponding functional forms are fitted to our first-principles data using 196 non-collinear states based on a Lebedev mesh for $\ell=2$ \cite{Lebedev1999}.

	\textbf{Adatoms on Au(111).} 
	To illustrate the different effects on the Gilbert damping, we start by exploring magnetic adatoms in the uniaxial symmetry of the Au(111) surface. For the adatoms no non-local effects can contribute to the Gilbert damping.  
	
	The Gilbert damping tensor of a single adatom without SOC has the form shown in relation to eq.~\eqref{eq:damping_on-site_single},
	\begin{align}
		\mathcal{G}_{i}^{0}=  \mathcal{S}_i \mathcal{I}  + \mathcal{A}_i \mathcal{E}(\vec{e}_i) \quad . \label{eq:gilbert_damping_C3v_adatom}
	\end{align}
	Note that SOC can induce additional anisotropies, as shown in eq.~\eqref{eq:second_order_dependencies_on_site_soc}.
	The most important ones for the case of a single adatom are $\{ \epsilon_{\alpha \beta \gamma} \hat{n}_{ij}^\gamma ,
	\hat{n}_{ij}^\alpha \hat{n}_{ji}^\beta \}$, which in the $C_{3v}$ symmetry result in
	\begin{align}
		\mathcal{G}_{i} =  \mathcal{G}_{i}^{0} + \mathcal{S}_i^\text{soc} \begin{pmatrix} 0 & 0 & 0 \\ 0 & 0 & 0 \\ 0 & 0 & 1 \end{pmatrix} + \mathcal{A}_i^\text{soc} \begin{pmatrix} 0 & 1 & 0 \\ -1 & 0 & 0 \\ 0 & 0 & 0 \end{pmatrix} \quad , \label{eq:gilbert_damping_C3v_adatom_soc}
	\end{align}
	since the sum of all SOC vectors points in the out-of-plane direction with $\hat{\vec{n}}_{ij} \rightarrow \vec{e}_z$.
	Thus, the Gilbert damping tensor of adatoms deposited on the Au(111) surface can be described by the four parameters shown in eqs.~\eqref{eq:gilbert_damping_C3v_adatom} and \eqref{eq:gilbert_damping_C3v_adatom_soc}, which are reported in Table~\ref{tab:damping_parameter_adatoms_DFT_Au111} for Cr, Mn, Fe and Co adatoms.
	Cr and Mn,  being nearly half-filled, are characterized by  a small damping-like contribution $\mathcal{S}_i$, while Fe and Co having states at the Fermi level show a significant damping of up to $\SI{0.47}{}$ in the case of Co.
	The antisymmetric part $\mathcal{A}_i$ of the Gilbert damping tensor results in an effective renormalization of the gyromagnetic ratio $\gamma$, as shown in relation to eq.~\eqref{eq:appB}, which using the full LLG equation, eq.~\eqref{eq:general_LLG}, and approximating $\vec{m}_i \cdot \frac{\dd \vec{m}_i}{\dd t} = 0$ is given by,
	\begin{align}
		\gamma^\text{renorm} = \gamma \, \frac{1}{1+\gamma (\vec{e}_i \cdot \vec{A}_i)} \quad , \label{eq:renormalized_gyromagnetic_ratio_c3v}
	\end{align}
	where $\vec{A}_i$ describes the vector $\vec{A}_i = \big( \mathcal{A}_i \, , \, \mathcal{A}_i \, , \,  \mathcal{A}_i+\mathcal{A}_i^\text{soc} 
	\big)$.
	For Cr and Fe there is a significant renormalization of the gyromagnetic ratio resulting in approximately $\SI{1.4}{}$.
	In contrast, Co shows only a weak renormalization with $\SI{1.9}{}$ being close to the gyromagnetic ratio of $\SI{2}{}$.
	The SOC effects are negigible for most adatoms except for Fe, which shows a small anisotropy in the renormalized gyromagentic ratio ($\approx \SI{10}{\percent}$) and a large anisotropy in the damping-like term of nearly $\SI{50}{\percent}$.
	
	\begin{table}[!t]
		\centering
		\begin{tabular}{l|rrrr}
			\hline 
			Damping   & \multirow{2}{*}{Cr / Au(111)} &  \multirow{2}{*}{Mn / Au(111)} &  \multirow{2}{*}{Fe / Au(111)} & \multirow{2}{*}{Co / Au(111)} \\
			parameters & & & \\ \hline
			$\mathcal{S}_i           $   &     \SI{0.083}{} & \SI{0.014}{} & \SI{ 0.242}{} & \SI{0.472}{} \\
			$\mathcal{A}_i           $   &     \SI{0.204}{} & \SI{0.100}{} & \SI{ 0.200}{} & \SI{0.024}{} \\
			$\mathcal{S}_i^\text{soc}$   &     \SI{0.000}{} & \SI{0.000}{} & \SI{ 0.116}{} & \SI{0.010}{} \\
			$\mathcal{A}_i^\text{soc}$   &     \SI{0.000}{} & \SI{0.000}{} & \SI{-0.022}{} & \SI{0.012}{} \\
			$\gamma^\text{renorm}_{x/y}$ &     \SI{1.42}{} & \SI{1.67}{} & \SI{ 1.43}{} & \SI{1.91}{} \\
			$\gamma^\text{renorm}_z$     &     \SI{1.42}{} & \SI{1.67}{} & \SI{1.48 }{} & \SI{1.87}{} \\
			\hline
		\end{tabular}
		\caption[Gilbert damping parameters of Cr, Mn, Fe and Co adatoms on Au(111).]{Gilbert damping parameters of Cr, Mn, Fe and Co adatoms deposited on the Au(111) surface as parametrized in eqs.~\eqref{eq:gilbert_damping_C3v_adatom} and \eqref{eq:gilbert_damping_C3v_adatom_soc}.
			The SOC field points in the $z$-direction due to the $C_{3v}$ symmetry.
			The renormalized gyromagnetic ratio $\gamma^\text{renorm}$ is calculated according to eqs.~\eqref{eq:renormalized_gyromagnetic_ratio_c3v} for an in-plane magnetic moment and an out-of-plane magnetic moment.
		}
		\label{tab:damping_parameter_adatoms_DFT_Au111}
	\end{table}
	
	\textbf{Dimers on Au(111).}
	In contrast to single adatoms, dimers can show non-local contributions and dependencies on the relative orientation of the magnetic moments carried by the atoms. 
	All quantities depending on the SOC vector are assumed to lie in the $y$-$z$-plane due to the mirror symmetry of the system. A sketch of the dimer and its nearest neighboring substrate atoms together with adatoms' local density of states are presented in Fig.~\ref{fig:figure3}. 
	
	The density of states originates mainly from the $d$-states of the dimer atoms.
	It can be seen that the dimers exhibit a much more complicated hybridization pattern than the Alexander-Anderson model.
	In addition the crystal field splits the different $d$-states resulting in a rich and high complexity than assumed in the model.
	However, the main features are comparable: For all dimers there is either a fully occupied majority channel (Mn, Fe, and Co) or a fully unoccupied minority channel (Cr).
	The other spin channel determines the magnetic moments of the dimer atoms $\{\SI{4.04}{}, \SI{4.48}{}, \SI{3.42}{}, \SI{2.20}{}\}\SI{}{\mub}$  for respectively Cr, Mn, Fe and Co. Using the maximal spin moment, which is according to Hund's rule $\SI{5}{\mub}$, the first-principles results can be converted to the single-orbital Alexander-Anderson model corresponding to approximately $m=\{0.81, 0.90, 0.68, 0.44 \}\SI{}{\mub}$ for the aforementioned sequence of atoms.
	Thus by this comparison, we expect large non-collinear contributions for Fe and Co, while Cr and Mn should show only weak non-local dependencies.
	\begin{table}[!t]
		\centering
		\begin{tabular}{l|rrrr}
			\hline
			Damping   & \multirow{2}{*}{Cr / Au(111)} &  \multirow{2}{*}{Mn / Au(111)} &  \multirow{2}{*}{Fe / Au(111)} & \multirow{2}{*}{Co / Au(111)} \\
			parameters & & & \\ \hline
			$\mathcal{S}_i                                          $   &           $ \SI{+0.0911}{}$    &    $\SI{+0.0210}{}$    &    $\SI{+0.2307}{}$    &    $\SI{+0.5235  }{}$  \\
			$\mathcal{S}_i^{ij,(1)                     }  $   &           $ \SI{+0.0376}{}$    &    $\SI{+0.0006}{}$    &    $\SI{-0.3924}{}$    &    $\SI{-0.2662  }{}$  \\
			$\mathcal{S}_i^{ij,(2)                     }  $   &           $ \SI{+0.0133}{}$    &    $\SI{-0.0006}{}$    &    $\SI{+0.3707}{}$    &    $\SI{+0.3119  }{}$  \\
			$\mathcal{A}_i^{                                   }  $   &           $ \SI{+0.2135}{}$    &    $\SI{+0.1158}{}$    &    $\SI{+0.1472}{}$    &    $\SI{+0.0915  }{}$  \\
			$\mathcal{A}_i^{ij                     }  $   &           $ \SI{+0.0521}{}$    &    $\SI{+0.0028}{}$    &    $\SI{-0.0710}{}$    &    $\SI{-0.0305  }{}$  \\
			$\mathcal{S}_{ij}^{                           }  $   &           $ \SI{-0.0356}{}$    &    $\SI{+0.0028}{}$    &    $\SI{+0.2932}{}$    &    $\SI{+0.0929  }{}$  \\
			$\mathcal{S}_{ij}^{\text{dot}                  }  $   &           $ \SI{-0.0344}{}$    &    $\SI{-0.0018}{}$    &    $\SI{-0.3396}{}$    &    $\SI{-0.4056  }{}$  \\
			$\mathcal{S}_{ij}^{\text{dot},(2)                  }  $   &           $ \SI{+0.0100}{}$    &    $\SI{+0.0001}{}$    &    $\SI{+0.1579}{}$    &    $\SI{+0.2468  }{}$  \\
			$\mathcal{A}_{ij}^{                          }  $   &           $ \SI{-0.0281}{}$    &    $\SI{-0.0044}{}$    &    $\SI{+0.0103}{}$    &    $\SI{+0.0011  }{}$  \\
			$\mathcal{A}_{ij}^{\text{dot}                }  $   &           $ \SI{-0.0175}{}$    &    $\SI{+0.0000}{}$    &    $\SI{-0.0234}{}$    &    $\SI{-0.0402  }{}$  \\
			$\mathcal{S}_{ij}^{\text{cross}             }  $   &           $ \SI{+0.0288}{}$    &    $\SI{+0.0002}{}$    &    $\SI{-0.2857}{}$    &    $\SI{-0.0895  }{}$  \\
			$\mathcal{S}_{ij}^{\text{ba}                         }  $   &           $ \SI{+0.0331}{}$    &    $\SI{+0.0036}{}$    &    $\SI{+0.2181}{}$    &    $\SI{+0.2651  }{}$  \\
			$\mathcal{S}_i^{\text{soc},ij,y                 }  $   &           $ \SI{+0.0034}{}$    &    $\SI{+0.0000}{}$    &    $\SI{+0.0143}{}$    &    $\SI{-0.0225  }{}$  \\
			$\mathcal{S}_i^{\text{soc},ij,z                 }  $   &           $ \SI{+0.0011}{}$    &    $\SI{+0.0000}{}$    &    $\SI{-0.0104}{}$    &    $\SI{+0.0156  }{}$  \\
			$\mathcal{A}_i^{\text{soc},ij,y                }  $   &           $ \SI{+0.0024}{}$    &    $\SI{-0.0001}{}$    &    $\SI{-0.0036}{}$    &    $\SI{+0.0022  }{}$  \\
			$\mathcal{A}_i^{\text{soc},ij,z                }  $   &           $ \SI{+0.0018}{}$    &    $\SI{-0.0005}{}$    &    $\SI{+0.0039}{}$    &    $\SI{-0.0144  }{}$  \\
			$\mathcal{S}_{ij}^{\text{soc},y              }  $   &           $ \SI{+0.0004}{}$    &    $\SI{+0.0001}{}$    &    $\SI{+0.0307}{}$    &    $\SI{+0.0159  }{}$  \\
			$\mathcal{S}_{ij}^{\text{soc},z              }  $   &           $ \SI{-0.0011}{}$    &    $\SI{+0.0000}{}$    &    $\SI{-0.0233}{}$    &    $\SI{+0.0206  }{}$  \\
			$\mathcal{S}_{ij}^{\text{ba,soc},y                   }  $   &           $ \SI{-0.0027}{}$    &    $\SI{+0.0000}{}$    &    $\SI{-0.0184}{}$    &    $\SI{-0.0270  }{}$  \\
			$\mathcal{S}_{ij}^{\text{ba,soc},z                   }  $   &           $ \SI{+0.0005}{}$    &    $\SI{-0.0001}{}$    &    $\SI{+0.0116}{}$    &    $\SI{-0.0411  }{}$  \\
			\hline
		\end{tabular}
		\caption{Damping parameters of Cr, Mn, Fe and Co dimers deposited on the Au(111) surface.
			The possible forms of the damping are taken from the analytic model.
			The SOC field  is assumed to lie in the $y$-$z$ plane and inverts under permutation of the two dimer atoms.
		}
		\label{tab:damping_parameter_DFT_Au111}
	\end{table}
	
	The obtained parametrization is given in Table \ref{tab:damping_parameter_DFT_Au111}.
	The Cr and Mn dimers show a weak or nearly no  directional dependence. 
	While the overall damping for both nanostructures is rather small, there is a significant correction to the gyromagnetic ratio. 
	
	In contrast, the Fe and Co dimers are characterized by a very strong directional dependence.
	Originating from the isotropic dependencies of the damping-like contributions, the damping of the Fe dimer can vary between $\SI{0.21}{}$ in the ferromagnetic state and $\SI{0.99}{}$ in the antiferromagnetic state.
	For the Co dimer the inter-site damping is even dominated by the bilinear and biquadratic term, while the constant damping is negligible.
	In total, there is a very good qualitative agreement between the expectations derived from studying the Alexander-Anderson model and the first-principles results.

	\section{Conclusions}
	In this article, we presented a comprehensive analysis of magnetization dynamics in non-collinear system with a special focus on the Gilbert damping tensor and its dependencies on the non-collinearity. 
	Using a perturbative expansion of the two-site Alexander-Anderson model, we could identify that both, the intra-site and the inter-site part of the Gilbert damping, depend isotropically on the  environment via the effective angle between the two magnetic moments, $\vec{e}_i \cdot \vec{e}_j$.
	SOC was identified as the source of a chiral contribution to the Gilbert damping, which similarly to the Dzyaloshinskii-Moriya and chiral biquadratic interactions depends linearly on the vector spin chirality, $\vec{e}_i \times \vec{e}_j$. We unveiled dependencies that are proportional to the three-spin scalar  chirality
	$\vec{e}_i \cdot(\vec{e}_j \times \vec{e}_k)$, i.e. to the chiral or topological moment, and to its square. 
	Using the Alexander-Anderson model, we investigated the importance of the different contributions in terms of their magnitude as function of the magnetization. 
	Using the prototypical test system of Cr, Mn, Fe and Co dimers deposited on the Au(111) surface, we extracted the effects of the non-collinearity on the Gilbert damping using time-dependent DFT.
	Overall, the first-principles results agree qualitatively well with the Alexander-Anderson model, showing no dependence for the nearly half-filled systems Cr and Mn and a strong dependence on the non-collinearity for Fe and Co having a half-filled minority spin-channel.
	The realistic systems indicate an even stronger dependence on the magnetic texture than the model with the used parameters.
	The Fe and the Co dimer show significant isotropic terms up to the biquadratic term, while the chiral contributions originating from SOC have only a weak impact on the total Gilbert damping.
	However, the chiral contributions can play the deciding role for systems which are degenerate in the isotropic terms, like e.g.~spin spirals of opposite chirality.
	
	We expect the dependencies of the Gilbert damping on the magnetic texture to have a significant and non-trivial impact on the spin dynamics of complex magnetic structures. Our findings are readily implementable in the  LLG model, which can trivially be amended with the angular dependencies provided in the manuscript. Utilizing multiscale mapping approaches, it is rather straightforward to generalize the presented forms for an implementation of the micromagnetic LLG and Thiele equations. 
	The impact of the different contributions to the Gilbert damping, e.g.~the vector (and/or scalar) chiral and the isotropic contributions, can be analyzed on the basis of  either free parameters or sophisticated parametrizations obtained from first principles as discussed in this manuscript.
	It remains to be explored how the newly found dependencies of the Gilbert damping affect the excitations and motion of a plethora of highly non-collinear magnetic quasi-particles such as magnetic skyrmions, bobbers, hopfions, domain walls and spin spirals.
	Future studies using atomistic spin dynamics simulations could shed some light on this aspect and help for the design of future devices based on spintronics.

	
	\begin{acknowledgments}
	This work was supported by the European Research Council (ERC) under the European Union's Horizon 2020 research and innovation program (ERC-consolidator grant 681405 -- DYNASORE) and from Deutsche For\-schungs\-gemeinschaft (DFG) through SPP 2137 ``Skyrmionics''  (Project LO 1659/8-1). 
	The authors gratefully acknowledge the computing time granted through JARA-HPC on the supercomputer JURECA at the Forschungszentrum J\"ulich \cite{jureca}.
	\end{acknowledgments}
	
	\section{Methods}
	

	\appendix
	\section{Analysis of the Gilbert damping tensor}\label{app:Gilbert_damping_tensor}
	The Gilbert damping tensor $\mathcal{G}$ can be decomposed into a symmetric part $\mathcal{S}$ and an anti-symmetric part $\mathcal{A}$,
	\begin{align}
		\mathcal{A} = \frac{\mathcal{G} - \mathcal{G}^\texttt{T}}{2} \qquad \text{and} \qquad \mathcal{S} = \frac{\mathcal{G} + \mathcal{G}^\texttt{T}}{2} \quad .
	\end{align}
	While the symmetric contribution can be referred to as the damping-like contribution including potential anisotropies, the anti-symmetric $\mathcal{A}$ typically renormalizes the gyromagnetic ratio as can be seen as follows: The three independent components of an anti-symmetric tensor can be encoded in a vector $A$ yielding
	\begin{align}
		\mathcal{A}_{\alpha \beta} = \epsilon_{\alpha \beta \gamma} A_\gamma \quad ,
	\end{align}
	where $\epsilon_{\alpha \beta \gamma}$ is the Levi-Cevita symbol.
	Inserting this into the LLG equation yields
	\begin{align}
		\frac{\ud \vec{m}_i}{\ud t} =& - \gamma \vec{m}_i \times \left( \vec{B}_i^\text{eff} + \sum_j \mathcal{A}_{ij} \frac{\ud \vec{m}_j}{\ud t} \right) \\
		\approx&  - \gamma \vec{m}_i \times \left( \vec{B}_i^\text{eff} - \gamma \sum_j \mathcal{A}_{ij}  \left(  \vec{m}_j \times \vec{B}_j^\text{eff}\right) \right) \quad .
	\end{align}
	The last term can be rewritten as
	\begin{align}
		\left(\vec{A}_{ij} \cdot \vec{m}_j\right) \vec{B}_j^\text{eff} - \left(\vec{A}_{ij} \cdot \vec{B}_j^\text{eff}\right) \vec{m}_j \quad .\label{eq:appB}
	\end{align}
	For the local contribution, $\mathcal{A}_{ii}$, the correction is $\parallel \vec{m}_i$ and $\parallel \vec{B}_i^\text{eff}$ yielding a renormalization of $\gamma \, \vec{m}_i \times \vec{B}_i^\text{eff}$ .
	However, the non-local parts of the anti-symmetric Gilbert damping tensor can be damping-like.

	\section{Relation between the LLG and the magnetic susceptibility}\label{app:relation_LLG_susc}
	The Fourier transform of the LLG equation is given by
	\begin{align}
		-i \omega \vec{m}_i = - \gamma \vec{m}_i \times \left( \vec{B}_i^\text{ext} - \sum_j J_{ij} \vec{m}_j - i \omega \sum_j \mathcal{G}_{ij}  \vec{m}_j \right) \quad .
	\end{align}
	Transforming this equation to the local frames of site $i$ and $j$ using the rotation matrices $\mathcal{R}_i$ and $\mathcal{R}_j$ yields
	\begin{align}
		\frac{i \omega}{\gamma M_i} \vec{m}_i^\text{loc} =  \frac{\vec{m}_i^\text{loc}}{M_i} \times \left( \mathcal{R}_i \vec{B}_i^\text{ext} - \sum_j \mathcal{R}_i J_{ij}  \mathcal{R}_j^\mathsf{T} \vec{m}_j ^\text{loc}- i \omega \sum_j \mathcal{R}_i \mathcal{G}_{ij} \mathcal{R}_j^\mathsf{T} \vec{m}_j^\text{loc}  \right) \quad ,
	\end{align}
	where $\vec{m}_i^\mathrm{loc}=\mathcal{R}_i \vec{m}_i$ and $\vec{m}_j^\mathrm{loc}=\mathcal{R}_j \vec{m}_j$. The rotation matrices are written as $\mathcal{R}(\vartheta_i,\varphi_i) = \cos(\vartheta_i/2) \sigma_0 + \iu \, \sin(\vartheta_i/2) \big( \sin(\varphi_i)  \, \sigma_x - \cos(\varphi_i) \, \sigma_y \big)$, with $(\vartheta_i,\varphi_i)$ being the polar and azimuthal angle pertaining to the moment $\vec{m}_i$. In the ground state the magnetic torque vanishes. 
	Thus, denoting $\vec{m}_i^\text{loc} = \left(m_i^x \ , \ m_i^y \ , \ M_i \right)$, where $m_i^{x/y}$ are perturbations to the ground states, yields for the ground state
	\begin{align}
		\begin{pmatrix}
			(\mathcal{R}_i \vec{B}_i^\text{ext} )^x - \sum_j (\mathcal{R}_i J_{ij} \mathcal{R}_j^\mathsf{T} M_j \vec{e}_z)^x 
			\\
			(\mathcal{R}_i \vec{B}_i^\text{ext} )^y - \sum_j (\mathcal{R}_i J_{ij} \mathcal{R}_j^\mathsf{T} M_j \vec{e}_z)^y 
			\\
			(\mathcal{R}_i \vec{B}_i^\text{ext} )^z - \sum_j (\mathcal{R}_i J_{ij} \mathcal{R}_j^\mathsf{T} M_j \vec{e}_z)^z 
		\end{pmatrix} 
		= 
		\begin{pmatrix} 0 \\ 0 \\ (\mathcal{R}_i \vec{B}_i^\text{eff})^z \end{pmatrix} \quad .
	\end{align}
	Linearizing the LLG and using the previous result and limiting our expansion to transveral excitations yield 
	\begin{align}
		\frac{\iu \omega }{\gamma M_i} m_i^x
		&=
		m_i^y \frac{(\mathcal{R}_i \vec{B}_i^\text{eff})^z}{M_i} - (\mathcal{R}_i \vec{B}_i^\text{ext})^y + \sum_j (\mathcal{R}_i J_{ij}  \mathcal{R}_j^\mathsf{T} \vec{m}_j^\text{loc})^y + \iu \omega \sum_j (\mathcal{R}_i \mathcal{G}_{ij}  \mathcal{R}_j^\mathsf{T} \vec{m}_j^\text{loc})^y
		\\
		\frac{\iu \omega }{\gamma M_i} m_i^y
		&=
		- m_i^x \frac{(\mathcal{R}_i \vec{B}_i^\text{eff})^z}{M_i} + (\mathcal{R}_i \vec{B}_i^\text{ext})^x - \sum_j (\mathcal{R}_i J_{ij}  \mathcal{R}_j^\mathsf{T} \vec{m}_j^\text{loc})^x - \iu \omega \sum_j (\mathcal{R}_i \mathcal{G}_{ij}  \mathcal{R}_j^\mathsf{T} \vec{m}_j^\text{loc})^x\quad ,
	\end{align}
	which in a compact form gives
	\begin{align}
		\sum_{\substack{j \\ \beta=x,y}} \left( \delta_{ij} \left(\delta_{\alpha \beta} \frac{(\mathcal{R}_i \vec{B}_i^\text{eff})^z}{M_i} + \epsilon_{\alpha \beta \mu} \frac{\iu \omega}{\gamma M_i} \right) + \sum_j (\mathcal{R}_i J_{ij}  \mathcal{R}_j^\mathsf{T} )^{\alpha \beta}+ \iu \omega \sum_j (\mathcal{R}_i \mathcal{G}_{ij}  \mathcal{R}_j^\mathsf{T} )^{\alpha \beta} \right) m_j^\beta = (\mathcal{R}_i \vec{B}_i^\text{ext})^\alpha \quad ,
	\end{align}
	and can be related to the inverse of the magnetic susceptibility 
	\begin{align}
		\sum_{\substack{j \\ \beta=x,y}} \chi^{-1}_{i\alpha,j\beta} (\omega) m_j^\beta = (\mathcal{R}_i \vec{B}_i^\text{ext})^\alpha \quad .
	\end{align}
	Thus, the magnetic susceptibility in the local frames of site $i$ and $j$ is given by
	\begin{align}
		\chi^{-1}_{i\alpha,j\beta} (\omega) = 
		\delta_{ij} \left(\delta_{\alpha \beta} \frac{(\mathcal{R}_i \vec{B}_i^\text{eff})^z}{M_i} + \epsilon_{\alpha \beta \mu} \frac{\iu \omega}{\gamma M_i} \right) + \sum_j (\mathcal{R}_i J_{ij}  \mathcal{R}_j^\mathsf{T} )^{\alpha \beta}+ \iu \omega \sum_j (\mathcal{R}_i \mathcal{G}_{ij}  \mathcal{R}_j^\mathsf{T} )^{\alpha \beta} 
	\end{align}

	\section{Alexander-Anderson model--more details} \label{app:Method_Alexander}
	We use a single orbital Alexander-Anderson model,
	\begin{align}
		\mathcal{H} = \sum_{ij} \left[ \delta_{ij} \left( E_d - \iu \Gamma - U_i \vec{m}_i \cdot \vec{\sigma} - \vec{B}_i \cdot \vec{\sigma} \right) - (1-\delta_{ij})\, t_{ij} \right] \quad , \label{eq:anderson_ham}
	\end{align}
	where $i$ and $j$ sum over all $n$-sites, $E_d$ is the energy of the localized orbitals, $\Gamma$ is the hybridization in the wide band limit, $U_i$ is the local interaction responsible for the formation of a magnetic moment, $\vec{m}_i$ is the magnetic moment of site $i$, $\vec{B}_i$ is an constraining or external magnetic field, $\sigma$ are the Pauli matrices, and $t_{ij}$ is the hopping parameter between site $i$ and $j$, which can be in general spin-dependent.
	SOC is added as spin-dependent hopping using a Rashba-like spin-momentum locking $t_{ij} = t \left(\cos \varphi_\text{R} \, \sigma_0 - \iu  \sin \varphi_\text{R} \, \vec{n}_{ij} \cdot \vec{\sigma}\right)$, where the spin-dependent hopping is characterized by its strength defined by $\varphi_\text{R}$ and its direction $\vec{n}_{ij}=-\vec{n}_{ji}$~\cite{Chaudhary2018}. 
	The eigenenergies and eigenstates of the model are given by,
	\begin{align} 
		\mathcal{H} \ket{n} = (E_n - \iu \Gamma) \ket{n} \quad .
	\end{align}
	The single particle Green function can be defined using the eigensystem,
	\begin{align}
		G(E + \iu \eta ) = \sum_{n} \frac{\ket{n}\bra{n}}{E-E_n + \iu \eta} \quad ,
	\end{align}
	where $\eta$ is an infinitesimal parameter defining the retarded ($\eta\rightarrow 0^+$) and advanced ($\eta\rightarrow 0^-$) Green function.
	The magnitude of the magnetic moment is determined self-consistently using
	\begin{align}
		\vec{m}_i = - \frac{1}{\pi}\, \im \Tr \int \dd E\,  \vec{\sigma} \, G_{ii}(E) \quad,
	\end{align}
	where $G_{ii}(E)$ is the local Green function of site $i$ depending on the magnetic moment.
	Using the magnetic torque exerted on the moment of site $i$,
	\begin{align}
		\frac{\dd \mathcal{H}}{\dd \hat{e}_i} = - \vec{m}_i B_i^\text{eff} \quad ,
	\end{align}
	magnetic constraining fields can be defined ensuring the stability of an arbitrary non-collinear configuration,
	\begin{align}
		\vec{B}^\text{constr} = - \mathcal{P}_\perp^\vec{m} \frac{\vec{m}_i}{|\vec{m}_i|} B_i^\text{eff}
		\quad \Rightarrow \quad 
		\mathcal{H}^\text{constr} = - \vec{B}^\text{constr} \cdot \vec{\sigma} \quad ,
	\end{align}
	where $\mathcal{P}_\perp^\vec{m}$ is the projection on the plane perpendicular to the moment $\vec{m}$.
	The constraining fields are added to the hamiltonian, eq.~\eqref{eq:anderson_ham}, and determined self-consistently.

	\section{Density functional theory--details\label{app:method_DFT}}
	
	The density functional theory calculations were performed with the Korringa-Kohn-Rostoker (KKR) Green function method. We assume the atomic sphere approximation for the the potential and include full charge density in the self-consistent scheme~\cite{papanikolaou_conceptual_2002}. 
	Exchange and correlation effects are treated in the local spin density approximation (LSDA) as parametrized by Vosko, Wilk and Nusair~\cite{vosko_accurate_1980}, and SOC is added to the scalar-relativistic approximation in a self-consistent fashion~\cite{Bauer2014}.
	We model the pristine surfaces utilizing a slab of 40 layers with the experimental lattice constant of Au assuming open boundary conditions in the stacking direction, and surrounded by two vacuum regions.
	No relaxation of the surface layer is considered, as it was shown to be negligible~\cite{blonski_density-functional_2009}.
	We use $450 \times 450$ $k$-points in the two-dimensional Brillouin zone, and the angular momentum expansions for the scattering problem are carried out up to $\ell_\text{max} = 3$.
	Each adatom is placed in the fcc-stacking position on the surface, using the embedding KKR method.
	Previously reported relaxations  towards the surface of $3d$ adatoms deposited on the Au(111) surface~\cite{Brinker2018} indicate a weak dependence of the relaxation on the chemical nature of the element.
	Therefore, we use a relaxation towards the surface of $\SI{20}{\percent}$ of the inter-layer distance for all the considered dimers.
	The embedding region consists of a spherical cluster around each magnetic adatom, including the nearest-neighbor surface atoms.
	The magnetic susceptibility is efficiently evaluated by utilizing a minimal \textit{spdf} basis built out of regular scattering solutions evaluated at two or more energies, by orthogonalizing their overlap matrix~\cite{dos_santos_dias_relativistic_2015}.
	We restrict ourselves to the transversal part of the susceptibility using only the adiabatic exchange-correlation kernel and treat the susceptibility in the local frames of sites $i$ and $j$.
	To investigate the dependence of the magnetic excitations on the non-collinarity of the system, we use all possible non-collinear states based on a Lebedev mesh for $\ell=2$~\cite{Lebedev1999}.

	\section*{References}
	\bibliography{Lib.bib}

\providecommand{\newblock}{}
\begin{thebibliography}{10}
\expandafter\ifx\csname url\endcsname\relax
  \def\url#1{{\tt #1}}\fi
\expandafter\ifx\csname urlprefix\endcsname\relax\def\urlprefix{URL }\fi
\providecommand{\eprint}[2][]{\url{#2}}

\bibitem{Fert2013}
Fert A, Cros V and Sampaio J 2013 {\em Nat. Nanotech.\/} {\bf 8} 152--156 ISSN
  1748-3387

\bibitem{Fert2017}
Fert A, Reyren N and Cros V 2017 {\em Nature Reviews Materials\/} {\bf 2} 17031
  ISSN 2058-8437

\bibitem{Bogdanov1994}
Bogdanov A and Hubert A 1994 {\em Journal of Magnetism and Magnetic
  Materials\/} {\bf 138} 255 -- 269 ISSN 0304-8853

\bibitem{Roessler2006}
R\"ossler U~K, Bogdanov A~N and Pfleiderer C 2006 {\em Nature\/} {\bf 442}
  797--801

\bibitem{Tai2018}
Tai J~S~B and Smalyukh I~I 2018 {\em Phys. Rev. Lett.\/} {\bf 121}(18) 187201
  \urlprefix\url{https://link.aps.org/doi/10.1103/PhysRevLett.121.187201}

\bibitem{Parkin2008}
Parkin S~S~P, Hayashi M and Thomas L 2008 {\em Science\/} {\bf 320} 190--194
  ISSN 0036-8075 (\textit{Preprint}
  \eprint{https://science.sciencemag.org/content/320/5873/190.full.pdf})
  \urlprefix\url{https://science.sciencemag.org/content/320/5873/190}

\bibitem{Landau1935}
Landau L~D and Lifshitz E 1935 {\em Phys. Z. Sowjet.\/} {\bf 8} 153

\bibitem{Gilbert2004}
Gilbert T~L 2004 {\em IEEE Transactions on Magnetics\/} {\bf 40} 3443

\bibitem{Eriksson2017}
Eriksson O, Bergman A, Bergqvist L and Hellsvik J 2017 {\em {Atomistic Spin
  Dynamics: Foundations and Applications}\/} ({Oxford University Press})

\bibitem{dos_santos_dias_relativistic_2015}
{dos Santos Dias} M, Schweflinghaus B, Bl{\"u}gel S and Lounis S 2015 {\em
  Physical Review B\/} {\bf 91} 075405

\bibitem{Lounis2015}
Lounis S, dos Santos~Dias M and Schweflinghaus B 2015 {\em Phys. Rev. B\/} {\bf
  91}(10) 104420
  \urlprefix\url{https://link.aps.org/doi/10.1103/PhysRevB.91.104420}

\bibitem{Guimaraes2017}
Guimar\~aes F~S~M, dos Santos~Dias M, Schweflinghaus B and Lounis S 2017 {\em
  Phys. Rev. B\/} {\bf 96}(14) 144401
  \urlprefix\url{https://link.aps.org/doi/10.1103/PhysRevB.96.144401}

\bibitem{Bhattacharjee2012}
Bhattacharjee S, Nordstr\"om L and Fransson J 2012 {\em Phys. Rev. Lett.\/}
  {\bf 108}(5) 057204
  \urlprefix\url{https://link.aps.org/doi/10.1103/PhysRevLett.108.057204}

\bibitem{Kambersky1970}
{Kambersk{\'y}} V 1970 {\em Canadian Journal of Physics\/} {\bf 48} 2906

\bibitem{Mizukami2002}
Mizukami S, Ando Y and Miyazaki T 2002 {\em Phys. Rev. B\/} {\bf 66}(10) 104413
  \urlprefix\url{https://link.aps.org/doi/10.1103/PhysRevB.66.104413}

\bibitem{Tserkovnyak2002}
Tserkovnyak Y, Brataas A and Bauer G~E~W 2002 {\em Phys. Rev. Lett.\/} {\bf
  88}(11) 117601
  \urlprefix\url{https://link.aps.org/doi/10.1103/PhysRevLett.88.117601}

\bibitem{Hayami2017}
Hayami S, Ozawa R and Motome Y 2017 {\em Phys. Rev. B\/} {\bf 95}(22) 224424
  \urlprefix\url{https://link.aps.org/doi/10.1103/PhysRevB.95.224424}

\bibitem{Brinker2019}
Brinker S, dos Santos~Dias M and Lounis S 2019 {\em New Journal of Physics\/}
  {\bf 21} 083015 \urlprefix\url{https://doi.org/10.1088%2F1367-2630%2Fab35c9}

\bibitem{Laszloffy2019}
L{\'a}szl{\'o}ffy A, R{\'o}zsa L, Palot{\'a}s K, Udvardi L and Szunyogh L 2019
  {\em Physical Review B\/} {\bf 99} 184430

\bibitem{Grytsiuk2020}
Grytsiuk S, Hanke J~P, Hoffmann M, Bouaziz J, Gomonay O, Bihlmayer G, Lounis S,
  Mokrousov Y and Bl{\"u}gel S 2020 {\em Nature Communications\/} {\bf 11} 1--7
  ISSN 2041-1723

\bibitem{Brinker2020}
Brinker S, dos Santos~Dias M and Lounis S 2020 {\em Phys. Rev. Research\/} {\bf
  2}(3) 033240
  \urlprefix\url{https://link.aps.org/doi/10.1103/PhysRevResearch.2.033240}

\bibitem{Lounis2020}
Lounis S 2020 {\em New Journal of Physics\/} {\bf 22} 103003

\bibitem{Dias2022}
dos Santos~Dias M, Brinker S, L\'aszl\'offy A, Ny\'ari B, Bl\"ugel S, Szunyogh
  L and Lounis S 2022 {\em Phys. Rev. B\/} {\bf 105}(2) 026402
  \urlprefix\url{https://link.aps.org/doi/10.1103/PhysRevB.105.026402}

\bibitem{Jue2015}
Ju{\'e} E, Safeer C~K, Drouard M, Lopez A, Balint P, Buda-Prejbeanu L, Boulle
  O, Auffret S, Schuhl A, Manchon A, Miron I~M and Gaudin G 2015 {\em Nature
  Materials\/} {\bf 15} 272 EP --
  \urlprefix\url{https://doi.org/10.1038/nmat4518}

\bibitem{Akosa2016}
Akosa C~A, Miron I~M, Gaudin G and Manchon A 2016 {\em Phys. Rev. B\/} {\bf
  93}(21) 214429
  \urlprefix\url{https://link.aps.org/doi/10.1103/PhysRevB.93.214429}

\bibitem{Freimuth2017}
Freimuth F, Bl\"ugel S and Mokrousov Y 2017 {\em Phys. Rev. B\/} {\bf 96}(10)
  104418 \urlprefix\url{https://link.aps.org/doi/10.1103/PhysRevB.96.104418}

\bibitem{Akosa2018}
Akosa C~A, Takeuchi A, Yuan Z and Tatara G 2018 {\em Phys. Rev. B\/} {\bf
  98}(18) 184424
  \urlprefix\url{https://link.aps.org/doi/10.1103/PhysRevB.98.184424}

\bibitem{Kim2018}
Kim K~W, Lee H~W, Lee K~J, Everschor-Sitte K, Gomonay O and Sinova J 2018 {\em
  Phys. Rev. B\/} {\bf 97}(10) 100402
  \urlprefix\url{https://link.aps.org/doi/10.1103/PhysRevB.97.100402}

\bibitem{Alexander1964}
Alexander S and Anderson P~W 1964 {\em Phys. Rev.\/} {\bf 133}(6A) A1594--A1603
  \urlprefix\url{https://link.aps.org/doi/10.1103/PhysRev.133.A1594}

\bibitem{Gross1985}
Gross E~K~U and Kohn W 1985 {\em Phys. Rev. Lett.\/} {\bf 55}(26) 2850--2852
  \urlprefix\url{https://link.aps.org/doi/10.1103/PhysRevLett.55.2850}

\bibitem{Lounis2010}
Lounis S, Costa A~T, Muniz R~B and Mills D~L 2010 {\em Phys. Rev. Lett.\/} {\bf
  105}(18) 187205
  \urlprefix\url{https://link.aps.org/doi/10.1103/PhysRevLett.105.187205}

\bibitem{Lounis2011}
Lounis S, Costa A~T, Muniz R~B and Mills D~L 2011 {\em Phys. Rev. B\/} {\bf
  83}(3) 035109
  \urlprefix\url{https://link.aps.org/doi/10.1103/PhysRevB.83.035109}

\bibitem{Guimaraes2019}
Guimar{\~{a}}es F~S~M, Suckert J~R, Chico J, Bouaziz J, dos Santos~Dias M and
  Lounis S 2019 {\em Journal of Physics: Condensed Matter\/} {\bf 31} 255802

\bibitem{Chaudhary2018}
Chaudhary G, Dias M~d~S, MacDonald A~H and Lounis S 2018 {\em Phys. Rev. B\/}
  {\bf 98}(13) 134404
  \urlprefix\url{https://link.aps.org/doi/10.1103/PhysRevB.98.134404}

\bibitem{Dias2016}
dos Santos~Dias M, Bouaziz J, Bouhassoune M, Bl\"ugel S and Lounis S 2016 {\em
  Nature Commun.\/} {\bf 7} 13613

\bibitem{Hanke2016}
Hanke J~P, Freimuth F, Nandy A~K, Zhang H, Bl\"ugel S and Mokrousov Y 2016 {\em
  Phys. Rev. B\/} {\bf 94}(12) 121114
  \urlprefix\url{https://link.aps.org/doi/10.1103/PhysRevB.94.121114}

\bibitem{Dias2017}
dos Santos~Dias M and Lounis S 2017 {\em Spintronics X\/} {\bf 10357} 136 --
  152 \urlprefix\url{https://doi.org/10.1117/12.2275305}

\bibitem{Lebedev1999}
Lebedev V~I and Laikov D 1999 {\em Doklady Mathematics\/} {\bf 59} 477--481

\bibitem{jureca}
{J\"{u}lich Supercomputing Centre} 2018 {\em Journal of large-scale research
  facilities\/} {\bf 4}
  \urlprefix\url{http://dx.doi.org/10.17815/jlsrf-4-121-1}

\bibitem{papanikolaou_conceptual_2002}
Papanikolaou N, Zeller R and Dederichs P~H 2002 {\em Journal of Physics:
  Condensed Matter\/} {\bf 14} 2799--2823

\bibitem{vosko_accurate_1980}
Vosko S~H, Wilk L and Nusair M 1980 {\em Canadian Journal of physics\/} {\bf
  58} 1200--1211

\bibitem{Bauer2014}
Bauer D~S~G 2014 {\em Development of a relativistic full-potential
  first-principles multiple scattering Green function method applied to complex
  magnetic textures of nano structures at surfaces\/} (Forschungszentrum
  J{\"u}lich J{\"u}lich)

\bibitem{blonski_density-functional_2009}
B{\l}o{\'n}ski P and Hafner J 2009 {\em Journal of Physics: Condensed Matter\/}
  {\bf 21} 426001 ISSN 0953-8984

\bibitem{Brinker2018}
Brinker S, dos Santos~Dias M and Lounis S 2018 {\em Phys. Rev. B\/} {\bf 98}(9)
  094428 \urlprefix\url{https://link.aps.org/doi/10.1103/PhysRevB.98.094428}

\end{thebibliography}
	
\end{document}